\documentclass[a4paper,fleqn,usenatbib]{mnras}

\usepackage{newtxtext,newtxmath}

\usepackage[T1]{fontenc}
\usepackage{ae,aecompl}

\usepackage{graphicx}	
\usepackage{amsmath}	

\def\lesssim{\mathrel{\hbox{\rlap{\hbox{\lower4pt\hbox{$\sim$}}}\hbox{$<$}}}}
\def\gtrsim{\mathrel{\hbox{\rlap{\hbox{\lower4pt\hbox{$\sim$}}}\hbox{$>$}}}}
\newcommand{\ltsima}{$\; \buildrel < \over \sim \;$}
\newcommand{\simlt}{\lower.5ex\hbox{\ltsima}}
\newcommand{\gtsima}{$\; \buildrel > \over \sim \;$}
\newcommand{\simgt}{\lower.5ex\hbox{\gtsima}}

\title[Sub-pc res. cosmological simulations ]{Sub-parsec resolution cosmological simulations of star-forming clumps at high redshift with feedback of individual stars}

\author[Calura et al.]
{F. Calura$^{1}$\thanks{E-mail: francesco.calura@inaf.it}, A. Lupi$^{2,3}$, J. Rosdahl$^{4}$, E. Vanzella$^{1}$, M. Meneghetti$^{1}$, P. Rosati$^{5}$, \newauthor
E. Vesperini$^{6}$, E. Lacchin$^{1,7}$, R. Pascale$^{1}$, R. Gilli$^{1}$
\\ ~ \\
$^{1}$INAF, Osservatorio Astronomico di Bologna, Via Gobetti 93/3, 40129 Bologna, Italy\\
$^{2}$Dipartimento di Fisica ``G. Occhialini'', Universit\`a degli Studi di Milano-Bicocca, Piazza della Scienza 3, I-20126 Milano, Italy\\
$^{3}$INFN, Sezione di Milano-Bicocca, Piazza della Scienza 3, I-20126 Milano, Italy\\
$^{4}$Univ Lyon, Univ Lyon1, Ens de Lyon, CNRS, Centre de Recherche Astrophysique de Lyon UMR5574, F-69230, Saint-Genis-Laval, France\\
$^{5}$UDipartimento di Fisica e Scienze della Terra, Universit\`a degli Studi di Ferrara, via Saragat 1, I-44122 Ferrara, Italy  \\
$^{6}$Department of Astronomy, Indiana University, Bloomington, IN 47401, USA\\
$^{7}$Dipartimento di Fisica e Astronomia dell’Universit\`a degli Studi di Bologna, via P. Gobetti 93/2, 40129 Bologna, Italy\\
}

\begin{document}

\label{firstpage}

\pagerange{\pageref{firstpage}--\pageref{lastpage}} \pubyear{2019}

\maketitle

\begin{abstract}
We introduce a new set of zoom-in cosmological simulations with sub-pc resolution, intended to model extremely faint,
highly magnified star-forming stellar clumps, detected at $z=6.14$ thanks to gravitational lensing. 
The simulations include feedback from individual massive stars (in both the pre-supernova and supernova phases), generated via stochastic,
direct sampling of the stellar initial mass function. 
We adopt a modified ’delayed cooling’ feedback scheme, specifically created to prevent artificial radiative loss of the energy injected by
individual stars in very dense gas $(n\sim 10^3-10^5$ cm$^{-3}$). 
The sites where star formation ignites are characterised by maximum densities of the order of $10^5$ cm$^{-3}$ and
gravitational pressures P$_{\rm grav}$/k $>  10^7$ K/cm$^3$,
corresponding to the values of the local, turbulent regions where the densest stellar aggregates form. 
The total stellar mass at $z=6.14$ is 3.4$\times10^7~\rm M_{\odot}$, in satisfactory agreement with the observed stellar mass of the observed systems. 
The most massive clumps have masses of $\sim 10^6~\rm M_{\odot}$ and half-mass sizes of $\sim 100$ pc. 
These sizes are larger than the observed ones, including also other samples of lensed high-redshift clumps, and imply an average density
one orders of magnitude lower than the observed one. 
In the size-mass plane, our clumps populate a sequence that is intermediate between the ones of observed high-redshift clumps and local dSph galaxies. 
\end{abstract}


\section{Introduction}
Deep high-redshift studies performed with the Hubble Space Telescope (HST) have made possible the detection of extremely faint objects,
with magnitudes as faint as $M_{\rm UV}~-14$ up to at redshifts $z=6–8$ (e.g. \citealt{van17a, liv17}). 
Some of these systems present very low stellar masses, down to a few $10^6~\rm M_{\odot}$
(e. g., \citealt{kar17}).  
However, the nature and classification of these systems are unclear: 
are they dwarf galaxies, HII galaxies \citep{ter16}, super star clusters, extremely compact star clusters or clumps? 
Answering to these questions requires estimates of a few basic physical quantities such as their stellar mass, star formation rate (SFR),
and size of these systems. In blank fields, these aspects cannot be investigated as faint star-forming galaxies
generally appear as very blue, point-like sources, due to the limited spatial resolution of current instruments
that can only probe their properties down to a few 100 pc \citep{elm17} at $z>2$. 
Gravitationally lensed fields, however, offer a unique opportunity to measure
physical properties such as light profiles and effective radii, even for very faint sources. Much progress has recently been driven
by deep observations of massive galaxy clusters, carried out in the context of large HST programmes, such as, e. g., the Hubble
Frontier Fields (HFF) survey \citep{lot17}. In these investigations, by exploiting gravitational lensing, clusters of galaxies are used as cosmic
telescopes to look deeply into the distant Universe. High-precision lensing models of galaxy clusters can be built using a large
number of multiply lensed sources, which generally span a large redshift range \citep{men17,ber19,ber20}. 
Thanks to strong lensing events generated by giant lenses located along the line of sight, distant sources can be magnified by large
factors (from ~ a few to 20, or even more, \citealt{van17a,bou21}), enabling us to analyse them with very high spatial resolution, large S/N and to
probe their structural parameters down to scales of a few tens of parsec (\citealt{van19}). 
In this framework, determining the source redshift is key. 
In this regard, remarkable progress has been made in recent years thanks
to the integral field spectrograph Multi-Unit
Spectroscopic Explorer (MUSE) on the Very Large Telescope \citep{bac10}, which has enabled the spectroscopic confirmation of hundreds of
multiple images in the redshift range $2\le z \le7$. 
These methods allowed us to determine absolute physical
quantities such as luminosities, sizes, stellar masses, and star formation rates of clustered star-forming regions up to $z>6$ \citep{van17a,van19}.
Recent deep MUSE observations of lensed fields enabled a census of tiny star-forming complexes 
(\citealt{van21a}), allowing us to peer into their internal structure, unveiling clumps with typical  
sizes of $\sim~100$ pc (\citealt{mes22}) and, in some cases, even breaking 
the clumps into star-forming complexes matching the scales of bound star clusters ($\le 20$ pc). 

To interpret appropriately such an impressive, new wealth of data, a solid understanding of the properties of
faint sources and their environment is needed.  
This requires improving our theoretical understanding of these systems and their evolutionary link with their local, more evolved counterparts. To this purpose, suitable instruments are hydrodynamic zoom-in simulations of the first galaxies,
that include an accurate description of several fundamental physical processes (e. g., \citealt{wet16,saw16,gra17,age20}).  
In this approach, a low resolution background realization of the large-scale structure surrounds a 
high-resolution region, centered on the halo of interest and which allows for a detailed implementation of the baryonic
physics, including star formation and stellar feedback (\citealt{vog20} and references therein).
However, in most cases, current state-of-the-art simulations reach a spatial resolution of the order of $\sim$10 pc and maximum
dark matter particle  mass resolution of
$\sim 10^3 \rm M_{\odot}$, therefore unsuited to probe the internal structure of compact clumps in early galaxies. 

In order to resolve the internal structure of the clumps, the spatial resolution 
must be at least of pc- (or sub-pc) scale. A sub-pc resolution is recommended not only to resolve the sub-structures, 
but also for a proper description of the turbulent star-forming gas. 
In fact, the filamentary structure of local star-forming regions probed by infrared observations 
have typical sizes of $\sim$0.1 pc \citep{arz11} and are thought to be formed by fragmentation of the parent molecular
cloud induced by turbulent shocks \citep{pad01}. Turbulence is a multi-scale process, in which large-scale 
motions of the interstellar medium (ISM) transfer energy down to small scales, i.e. of the order of the size of the filaments (e. g., \citealt{bou10}).
As a consequence of this, a proper description of the star-forming gas requires simulations spanning
a large dynamic range in physical scale \citep{ren13}. 
The capability to resolve the star-forming gas down to sub-pc scales, in turn, requires a proper modelling of the properties
of the newly formed stars. 
In cosmological simulations, the stellar component is generaly modelled by means of particles, aimed to represent stellar populations.  
In the assumption that a stellar particle samples the entire stellar 
initial mass function  (IMF), their properties such as the mass, energy and metals restored into the interstellar medium (also known as
'feedback'), are computed via IMF-averaged prescriptions. This approximation breaks down at high spatial (or mass-) resolution,
where stochastic variations in stellar populations become non-negligible (\citealt{sor17,smi21}),  
or when the available gas for star formation in a cell amounts to a few 10 $\rm M_{\odot}$ only, i.e. in cases in which the entire IMF is not sampled.
Such cases require a detailed modelling of individual stars. However, 
since resolving the collapse of individual stars would require a spatial resolution currently unfeasible
in cosmological simulations for stellar-evolution timescales (ranging from a few Myr to a Hubble time),
an alternative, valid approximation is the stochastic IMF-sampling (\citealt{sor17,wal20}). 
In order to be able to resolve the formation of the star-forming clumps, from their birth and in a fully cosmological context,
in this paper we present cosmological hydrodynamic simulations with sub-pc resolution and feedback of individual stars, modelled
through Poisson sampling of the IMF.
These simulations are the first within a new project aimed at SImulating the Environment where Globular clusters Emerged (SIEGE). 
This work is the first of a series, 
where we present the physical ingredients of the model and the implementation of the most fundamental 
baryonic physics recipes. 
We also describe our first results on the structural properties of the simulated systems, 
comparing them to the observational properties of the star-forming clumps detected at high redshift.

To our knowledge, ours represents one of the first attempts to model the feedback of single stars 
at sub-pc resolution, in a cosmological simulation and with a grid code. 
Previous attempts to model the feedback of individual massive stars in a non-cosmological framework include the 
isolated mergers studied by \cite{lah20}, dwarf galaxies \citep{eme19,gut21} and a Milky Way-like galaxy \citep{and20}. 
Performing cosmological simulations in this extremely high resolution regime has recently become feasible, 
although the attempts performed so far are still very rare (\citealt{gut22} and references therein). \\
The paper is organised as follows. In Sect. 2, the setup of simulations and the basic model ingredients
are presented.
In Sect. 3, we present and discuss our results and in Sect. 4, we draw our conclusions.
Throughout this paper we adopt a flat cosmological model with matter density parameter $\Omega_{\rm m}=0.276$ and Hubble
constant $H_{\rm 0}=70.3$ km s$^{-1}$ Mpc$^{-1}$ \citep{sha14, omo19}.

\begin{figure*}
	\includegraphics[width=12.5cm,height=8.9cm]{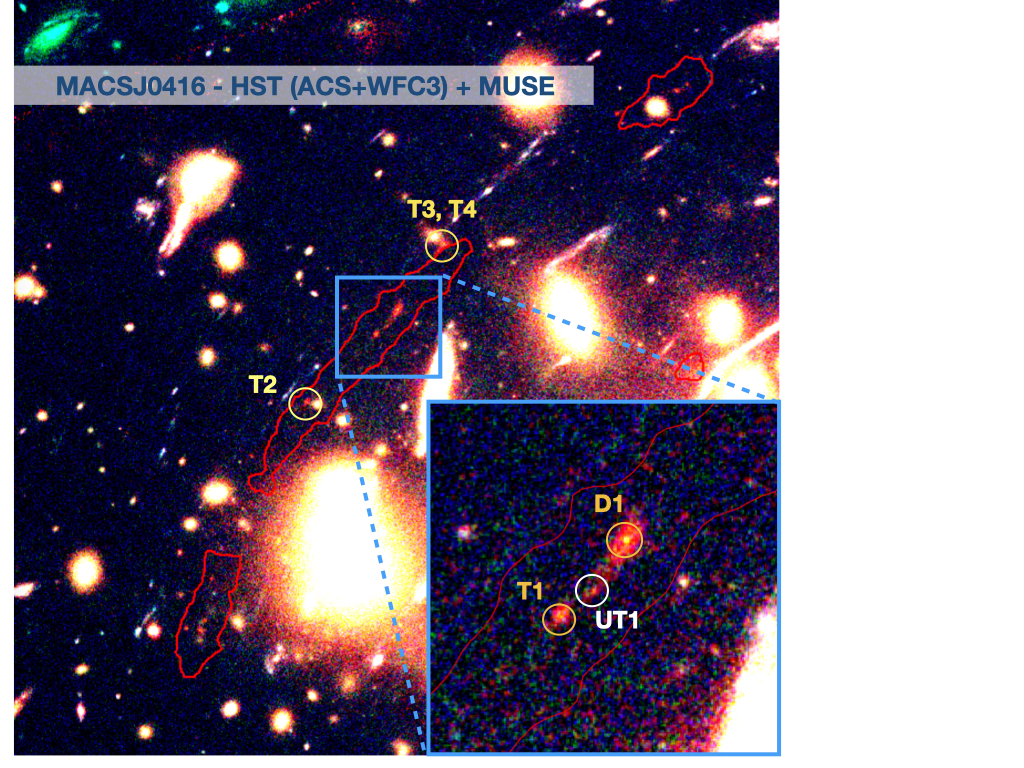}
\caption{Colour-composite image of the field containing the lensed,
spectroscopically confirmed star forming complexes at $z=6.14$ 
magnified by the galaxy cluster MACS J0416 \citep{van19}. The extended Lyman$-\alpha$ arcs 
detected with MUSE at $2-\sigma$ are shown by the red contours. 
The systems D1 and T1 are shown in the bottom-right inset.
A structure is visible between D1 and T1, including a very faint
star-forming knot, dubbed UT1, indicated with the white circle.  
Other detected sources include T2, T3, T4 and others \citep{van21b}, with intrinsic apparent
magnitudes in the HST F105W band between 30 and 32. }
    \label{D1T1}
\end{figure*} 
\section{Simulation Setup}
The simulation is aimed at modelling a system with features similar to an  
extended star-forming complex at $z=6.14$ strongly magnified by the galaxy cluster MACS J0416.1-2403, 
which includes several clumps, distributed across a wide region (\citealt{van19,cal21}).
The accurate model of the gravitational lens magnifying these systems was tested very carefully,
and was able to accurately account for the positions of a rich set of spectroscopically confirmed
multiple images in the redshift range $3 \le z \le 6.5$ (\citealt{ber20}). 
The UV emission is generated by extremely faint systems, characterised by intrinsic magnitudes $m_{\rm 1500 \AA} = 28 - 33$.
The total stellar mass of the star-forming complex is of a few 
$10^7~\rm M_{\odot}$. The main components of the complex are D1 and T1, shown in Fig. ~\ref{D1T1}. 
Other individual star-forming knots are present  
with intrinsic UV magnitudes between 30 and 32, 
intrinsic sizes between 10 and 50 pc and ages of 1-10
Myr, obtained from the analysis of their spectral energy distribution.
The sources are located in a region of varying, high magnification (typically $\mu > 15$). 
D1 has a stellar mass of $2.2\times 10^7~\rm M_{\odot}$, intrinsic (de-lensed) UVB magnitude 
29.60 and size of 44 pc \citep{van19}. 
T1 has an intrinsic UV magnitude of 31.3 and is one of the
faintest spectroscopically confirmed star-forming objects ever
identified at high redshift. Its stellar mass is $~2 \times 10^6~\rm M_{\odot}$ and its
size is $<30$ pc \citep{van17a}. 
In the available images obtained with HST, D1 appears considerably more extended than T1 and 
is characterised by a nucleated star-forming region surrounded by a diffuse component. 
D1 and T1 are separated by
1.7'', corresponding to a physical distance of $\sim~500$ pc assuming a magnification $\mu=20$ \citep{van17a}.   
The unique D1+T1 system provides deep insight into an extended star-forming region detected at the peak of reionization,  
in which a high magnification allows us to probe the complexity of its substructures.  

We model the evolution of such system by means of a zoom-in cosmological simulation 
performed with the adaptive mesh refinement hydrodynamic code RAMSES \citep{tey02}.
Gas evolution is computed using a second-order Godunov scheme for the Euler equations.
The Euler equations are solved using the second-order MUSCL scheme with an HLLC
Riemann solver, and the Poisson equation is computed using a
multi-grid method \citep{gui11}. Collisionless stellar particles are allowed to form, and their trajectories are
computed by means of a Particle-Mesh solver, as are the trajectories of dark matter (DM) particles,
with maximum mass resolution of 200 $\rm M_{\odot}$.  
 In the grid-based N-body scheme used by RAMSES, i. e.
the standard particle mesh method with adaptive mesh refinement, 
at each grid level, the gravitational softening length is equal to the local grid size,
without any distinction between stars and DM. 

In our simulations, we aim at pushing the spatial resolution to the sub-pc scale. 
The initial baryonic grid maximum spatial resolution is 1.2 $h^{-1}$ kpc comoving,
corresponding to maximum refinement level 12 and computed by means of a variable-based refinement strategy.
We allow for 9 additional refinement levels (with the cell width decreasing by half for each additional level), 
with a Lagrangian mass threshold-based criterion.

A cell is refined if it is not at the maximum level and its mass exceeds $8 m_{sph}$, 
 where $m_{sph}=\Omega_b/\Omega_m 2^{-3l}=32$M$_{\odot}$, in which $\Omega_b=0.045$ is the baryon density parameter
 and $l=12$ is the maximum grid level of the initial conditions.
This allows us to reach a maximum physical resolution of 0.3 $h^{-1}$ pc in the densest regions at $z=6.14$, corresponding to maximum refiment level lmax=21.
To check the numerical converge, we also run another lower resolution simulation with maximum refinement level 19,
corresponding to a minimum cell width of 1.9 pc at $z=6.14$. 
Besides star formation, our simulations include radiative cooling and feedback from individual asymptotic giant branch stars (AGB) and
massive stars (MS) in the forms of stellar winds and supernova explosions.
The high resolution of our simulation requires the feedback of single stars (and not of stellar particles, i. e. particles representing entire stellar populations)
to be properly taken into account. To this purpose, our implementation of star formation is designed in order to 
account for the formation of single stars, in an appropriate range of mass. 
In the following, we will describe the methods used to perform our task.

In our simulations, radiative cooling is modelled through the RAMSES native implementation,
i. e. as due to hydrogen, helium and metals (e. g., \citealt{few14}). 
The native cooling implementation of RAMSES is based on equilibrium-thermochemistry, and is generally
suited to study the dense star-forming gas \citep{age13,fic22}. Cooling and
heating rates of the gas are functions of temperature, density, redshift, metallicity, and the abundances
of each primordial ion species, $n_{\rm HI} , n_{\rm HII} , n_{\rm HeI}, n_{\rm HeII}$, and $n_{e}$. In RAMSES, photoionization equilibrium
is assumed, in which the primordial ion abundances are functions of temperature, density, and redshift,
calculated with a simple iterative process that involves equating the rates of photo-ionization, collisional
ionization and recombination. These rates are pre-computed and stored in tables every coarse time-step
in different bins of temperature and hydrogen number density $n_H=X~ \rho/m_H$, where $\rho$ is the gas density,
$m_H$ is proton mass and $X$ is the hydrogen mass fraction, set to the constant value of 0.76 (for further details, see \citealt{ros12}). 
A homogeneous, redshift-dependent ionizing UV background is also assumed \citep{haa96}. 

In our simulations, star formation begins at $z \simeq 16$, in good agreement 
with other models that include more sophisticated thermal modelling (e. g., \citealt{mai09,ros18}).  
After the formation of the first stars, the ISM is quickly polluted 
with metals and the cooling becomes dominated by the metal cooling rates, in general much higher than
primordial H2 cooling rates(e. g., \citealt{ros12}, \citealt{age13}). 
We assume a temperature floor of $T_{\rm floor} = 100$ K and
an intially metal-free gas.

In our simulations, we do not consider the effects of pop III stars.
In some works, their contribution to primordial metal enrichment is modeled directly \citep{wis12,gut22}. 
Alternately, an homogeneous, initial metallicity floor (typically of metallicity $10^{-4}~Z_{\odot}$)
is assumed to compensate for the lack of their modelling. 
The early feedback from PopIII stars is an interesting subject, but due to the large uncertainty in
their initial mass function and metallicity transition, their inclusion will imply the study of an additional set of free
parameters, which we prefer to skip in this presentation paper of our new simulations.

\subsection{Initial Conditions}
We need to find a suitable DM halo to host a
system with a comparable stellar mass as the D1+T1 system at $z=6.14$.
Very few constraints exist on the stellar-to-halo mass relation for low-mass halos $(M<10^{11}$M$_{\odot})$ at these redshifts,
where the extant studies find significantly different results \citep{sun16,rod17,beh19,ma19}.  

With the awareness that our DM halo lies across the edge of the existing relations at these redshifts, 
a stellar mass of a few $10^7$ M$_{\odot}$ 
is expected to correspond to a DM mass of $\sim$ a few $10^{10}~\rm M_{\odot}$ (\citealt{beh19} and references therein).   
Such halo mass is not particularly large and rather frequent even at high redshift.
For this reason, a small box is enough to guarantee the presence of at least
one of these objects within the simulated domain.

It is worth stressing that the choice of a small volume is key for reaching a sub-pc resolution
at the final redshift of our simulation. 
A comoving volume of 5 Mpc h$^{-1}$ is thus ideal for accomplishing our task. 
The initial conditions (ICs) are generated at z = 100 by means of the MUSIC software \citep{hah11}. 
To define the zoom-in region and increase resolution up to the desidered level, we follow 
the approach described in \cite{fia17} and \cite{lup19}, which consists of various steps.
\begin{enumerate}
\item First, the entire periodic box is covered with 64$^3$ root cells and a coarse-grid, 
N-body, DM-only simulation is run to $z=6.14$. 
\item At this stage, all the halos within the box are identified by means of the HOP halo finder \citep{eis98}.
The main algorithm behind HOP is basically designed to connect each particle to other nearby particles in the direction
of increasing density. 
This is performed by assigning a density to every particle, using an adaptive kernel with
a length scale set by the distance to a set of nearest analogues. 
After running HOP, we obtain a list of suitable halos and focus on the ones with more than 100 particles.  
After that, we select an isolated halo located in the central regions of the box with suitable
features, i.e. mass of $2-5~\times~10^{10}~\rm M_{\odot}$ and not in the process of merging. 
We also ensure that no other DM halo with comparable mass is present within a sphere with radius of $3$ times $R_{\rm vir}$. 
Here, the quantity $R_{\rm vir}$ is the virial radius. 
This quantity is defined as the physical size of a region within
which the matter density is $\Delta(z)~\rho_{\rm c}(z)$, where $\rho_{\rm c}(z)$ is the critical density of the Universe and
$\Delta(z)$ is the $z-$dependent virial overdensity, obtained by means of the fitting formula of Bryan \& Norman (1998;
see also \citealt{bar01}).
The virial mass of the halo is thus $M_{\rm vir} = \frac{4}{3}~\pi~\Delta(z)~\rho_{\rm c}(z)~R_{\rm vir}^3$ and, 
for our selected halo, it is  $\sim 4\times10^{10}~\rm M_{\odot}$ at $z=6.14$.

All the DM particles belonging to the selected halo are flagged and traced  back in time to the ICs. 
Subsequently, \item we increase the resolution of the ICs, adding two refinement levels to the Lagrangian region surrounding
the previously flagged DM particles, 
and we rerun the DM-only simulation. We then repeate the halo finding procedure (ii) and step (iii), adding 
two more refinement levels in a convex hull region around the flagged DM particles and re-running the DM-only simulation. 
As a result, we initially add six additional levels of refinement above the starting base cube in the zoom-in region.
We performed a further full, high-resolution DM-only test in order to ensure that the procedure described above
allows us to mantain a fraction of contaminating particles below 0.1 percent. 
The initial number of DM particles in our simulations is 2 $\times$ 10$^8$.
After this last test, the ICs are recomputed with the final inclusion of baryons. 
\end{enumerate}

\subsection{Star formation and IMF sampling} 

We assume that star formation can occur in cells with gas temperature $T<2 \times 10^4$ K. 
The gas, characterised by density $\rho$, can be converted into stars with density ${\rho}_{*}$ according to:
\begin{equation}
  \dot{\rho}_{*} = \frac{\rho}{t_*}, 
\end{equation}
i.e. according to the \cite{sch59} law. 
The star formation timescale $t_*$ is proportional to the local free-fall time $t_{\rm ff}$, and computed as
\begin{equation}
  t_{*}=  t_{\rm ff}/\epsilon_{\rm ff},  
\end{equation}
where $t_{\rm ff}=\sqrt{3~\pi/32~G \rho}$ and $\epsilon_{\rm ff}=0.1$ is the star formation efficiency per free-fall time,
consistent with direct observational estimates of this quantity in the most massive, local molecular clouds (\citealt{age15,mur11}). 
Moreover, by means of cosmological hydrodynamic simulations of the progenitors of Milky Way-sized galaxies,  
\cite{age15} showed how such a high star formation efficiency provides a good match to several
observables at differents redshifts, including the derived star formation histories, the stellar mass-gas metallicity relation
and its evolution, the Kennicutt-Schmidt relation and  the stellar mass - DM halo relation. 
More recently, \cite{gri19} demonstrated how high efficiencies $\epsilon_{\rm ff}=0.1$ on scales of parsecs
provides a close match to the observed efficiencies on scales of individual molecular clouds.

 We allow for no more than $90 \%$ of the gas in the cell to be turned into stars. This condition implies a minimum density
threshold for star formation (\citealt{yag22}):
\begin{equation}
  \rho_{\rm thr}= \frac{m_{\rm *}}{0.9 (\Delta x)^3},   
\end{equation}
where the quantity $m_{*}$ represents the base star particle mass. In RAMSES, one sets the value of this quantity
and the total particle mass in a 
cell will be a multiple N of $m_{*}$, determined by means of a Poisson sampling method as described in various
papers (e. g., see eq. 6 of \citealt{ras06}). 
In our case, we assume $m_{*}=m_{\rm sph}$, whereas $\Delta x$ is the width of the maximum resolution cell.
This corresponds to a typical value for the particle density threshold of  $\sim 10^5$ cm$^{-3}$.

To justify our adopted star fomation threshold, a note on artificial fragmentation is in order. 
Together with the assumption that star formation occurs at $T < 2\times 10^4$ K, this SF threshold
creates conditions in which the Jeans length is almost always marginally resolved.
This is visible from the phase diagrams shown in Fig.~\ref{fig_phase}, where we quantify the number of cells with under-resolved Jeans length. 

The problem of artificial fragmentation is common to various works in which very high densities and cold
gas temperatures are achieved. Sometimes, to tackle this issue, an artificial pressure floor is adopted (e. g, \citealt{ren13}).
One problem with the assumption of a pressure floor is that it may create further artificial effects (see \citealt{ble14}), 
as it effectively represents an additional source of feedback. In light of these
facts, as our model is capable of forming stars fast enough at high densities and to contain reasobably 
the effects of unresolved Jeans lengths, here we choose not to apply any density-dependent pressure floor. 

A growing body of works supports the need for individual stellar feedback in high resolution (i. e. pc-scale or higher) simulations
(\citealt{hu17,eme19,ste20}). 
Other recent works have shown that in high-resolution simulations, the modelling of the stellar component as 
Initial Mass Function (IMF)-averaged properties may result in an incorrect accounting of the feedback budget (e. g.,
\citealt{smi21}). 
As a general, conservative rule, \cite{smi21} suggest that the creation of individual stars via IMF-sampling
is preferred for particle masses below a few 100 $\rm M_{\odot}$. 
More generally, generating individual stars is preferred when the quantity of gas available for
star formation in a cell is sufficient for a few stars only. 
Such condition may be frequent at sub-pc resolution and for gas densities of $\sim 10^3-10^5 cm^{-3}$, i. e. for values
typical of dense star-forming regions such as molecolar cloud cores. 
The high resolution of our simulations requires therefore that we implement dedicated recipes for modelling the formation of individual MS. 
To this purpose, we follow the prescriptions described in \cite{sor17} and in \cite{and20}. 
We adopt a Kroupa (2001) IMF $\xi_{\rm K01}(m)$, defined as: 

\begin{equation}
\xi_{\rm K01}(m) =  \Big\{ \begin{array}{l l} 
                                      A \cdot m^{-0.3} & 
			     \qquad {\mathrm{if}} \; m < 0.5 \, \rm M_\odot \\
			              B \cdot m^{-1.3} &
			     \qquad {\mathrm{if}} \; m \ge 0.5 \, \rm M_\odot, \\
                                      \end{array}                                      
                                      \end{equation}
where the normalization constants $A, B$ are computed by imposing continuity and that
\begin{equation}
\int_{0.1 \rm M_{\odot}}^{100\rm M_{\odot}} \xi_{\rm K01}(m) = 1.
\end{equation} 
We decompose the stellar mass spectrum into $N$ finite intervals. 
In each mass interval, a mass fraction $f_i$ is defined as
\begin{equation}
\sum_{i=1}^{N} f_{i} = 1. 
\end{equation}

In the $i-$th mass interval, the number of individial stars $n_i$ can be determined 
by sampling from a Poisson distribution, characterised by a probability $P_i$ given by
\begin{equation}
P_i(n_i)={\lambda_i^{n_i} \over n_i !} \exp({-\lambda_i}) \,
\label{poisson_law}
\end{equation} 
where the mean value $\lambda_i$ is calculated as: 
\begin{equation}
\lambda_i= f_{i} {M \over m_i}, 
\label{poisson_param}
\end{equation}
where $M$ is the total mass available for star formation (re-corrected in the case it exceeds 90 \% of the gas mass)  
and $m_i$ the average stellar mass in the $i-$th bin. 
In this procedure, we Poisson-sample the number of particles to be made in each bin,
except the lowest bin, in which at most only one particle is spawn, that collects all
the lowest-mass stars together.
It is worth pointing out that, in its original presenation, the scheme of \cite{sor17} is useful to assign
a population of stellar masses to one star or sink particle, whereas in our case we spawn  
individal particles, each representing a single star.  

In the formalism of \cite{sor17}, the sampled stellar mass $M_{\rm sampled}$
\begin{equation} 
M_{\rm sampled} =   \sum_{i=1}^{N} n_i  m_i 
\end{equation}
converges towards $M$ for large numbers of stars, due to stochastic sampling.
As noted by \cite{and20}, as the sampling requires one 
random number for each bin, this method is optimal in simulations since
the computational expense is determined by the chosen number of mass bins. 
This means that the individual stars in each mass bin will all have the same
mass, but this is acceptable since, for large number of stars, the entire mass range will be correctly sampled \citep{sor17}.
Another important advantage is that the bin sizes can be chosen arbitrarily. In this way, in order
to limit the total particle number, this allows us to group all the stars below a
certain mass threshold in one single bin, while stars above this value are sampled individually.
In our case, we adopt $N=12$ bins and assume that all the stars in the lowest mass bin (characterised by $m_i=1.55~M_\odot$) are not formed as individual stars but
grouped into single particles. The reason for this choice is that this mass value corresponds to a Main-Sequence turn-off $>1$ Gyr, larger
than the cosmic age of the Universe at the end of our simulation ($t_{\rm end}=0.9$ Gyr). 
This choice is sufficient to recover approximately the correct fraction of massive stars (i.e. with initial mass $>8~M_{\odot}$). 

One problem with direct Poisson IMF sampling in simulation cells is that occasionaly, a stellar mass larger than
the effectively available particle mass may be generated. To avoid this problem, we have enforced a check
in the stellar mass created at each sampling, in which the excess with respect to the available gas
mass has been re-subtracted from the stellar budget in order to conserve mass. 
One important choice in this case concerns the stellar mass bin in which 
the mass in excess has to be subtracted.  In our case, we start creating individual stars from
the highest-mass bin and, in each bin, we check if the available mass is overflown. If this
is the case, we correct the mass created in that bin by subtracting the mass in excess and stop sampling in the cell. 
This was performed in order to not underestimate the number of MS, which may have important consequences on the
estimated feedback budget. We have re-checked a posteriori that the sampled amount of low-mass stars was not too small
compared to the IMF-calculated expectations, and verified that it is consistent to within less than 10 percent.

\begin{figure*}\includegraphics[width=18.cm,height=12.cm]{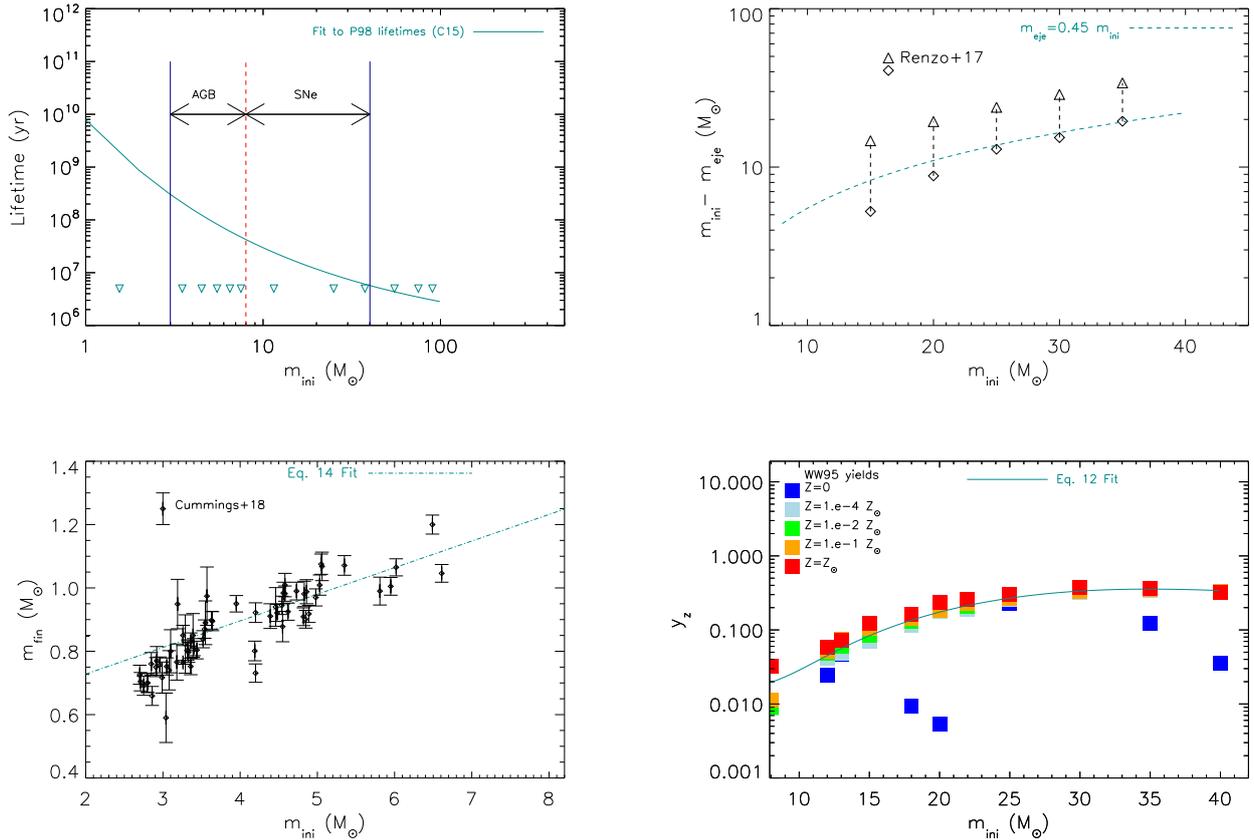}
\caption{Stellar evolution prescriptions adopted in the present work.
The top-left panel shows the stellar lifetimes as a function of the initial mass (dark-cyan solid line) 
expressed by an analytical fit of the \citet{por98} stellar lifetimes (\citealt{cai15}, see Eq.~\ref{eq_life}). 
The range in which individual stars are created 
is between 3 $\rm M_{\odot}$ and 100 $\rm M_{\odot}$.
We assume that stars with mass $>40 \rm M_{\odot}$ do not contribute to stellar feedback. 
The inverted dark-cyan triangles show the centers of the 12 mass bins selected for our IMF sampling. 
The vertical red dashed line marks the separation mass between
AGB and the progenitors of core-collapse SNe, i.e. the massive stars, with initial  mass  $8~\rm M_{\odot}< m_{\rm ini} <40~\rm M_{\odot}$. 
The vertical solid blue lines enclose the AGB+SNe mass range. 
The top-right panel shows the prescriptions regarding the mass ejected by stellar winds during the pre-SN phase. 
This is expressed by the analytical relation used for the adopted mass remnant 
at the end of the pre-SN phase (dashed dark cyan line), $m_{\rm eje}= 0.45 m_{\rm ini}$ (see Sect.~\ref{sec_fee}). 
The open triangles and diamonds plotted at 5 different masses enclose the range of final pre-SN mass values computed by
means of stellar evolution models for massive stars \citep{ren17}. The bottom-left panel shows the adopted analytical fit
(dark cyan dash-dotted line) to the final-to-initial stellar mass relation derived by \citet{cum18} (small black open diamonds, plotted with the error bars),
expressed by Eq.~\ref{eq_mfin}.  
Finally, in the bottom-right panel we show the analytical fit of Eq. \ref{eq_yield} used for the fractional
amount of heavy elements ejected by massive stars (dark-cyan dashed line) obtained from the \citet{woo95} yields at different
metallicities (coloured solid squares).}  
\label{fig_fee}
\end{figure*} 

\subsection{Stellar feedback}
\label{sec_fee}
In our simulations, stars are divided in three categories: MS (with mass $m \ge 8 M_\odot$),
intermediate mass stars (in the mass range  $3\le m < 8 M_\odot$) and low-mass stars (with $m < 3 M_\odot$). 
In our code we model the feedback of each individual massive and intermediate mass star,
whereas low-mass stars are collected into standard stellar particles. This choice is performed 
both in order to avoid dealing with a useless and excessive number of stars and also because low-mass stars
do not contribute to stellar feedback, as they live longer than the age of the Universe at the final time of the simulation.
In the remainder of the paper, we will distinguish between the 'individual star' particles, born with mass
$>3 M_\odot$ and 'stellar particles', containing all the stars with mass  $<3 M_\odot$. 
Fig. ~\ref{fig_fee} summarises all our prescriptions regarding the evolution of individual stars. 
The adopted stellar lifetimes are from \cite{por98}; here we use the analytical fit from \cite{cai15}, expressed as:
\begin{equation}
\tau (\rm yr) = 10^{C_1 (m_{\rm ini}/\rm M_{\odot})^{-C2} + C_3},
\label{eq_life}
\end{equation}
with $C_1=4.19$, $C_2=0.37$, $C_3=5.71$.

\subsubsection{Massive stars}
For MS, we consider two types of feedback: (i) pre-supernova and (ii) Type-II SN feedback. 
We also assume that only stars with mass $<40 M_\odot$ contribute to both feedback and metal enrichment, 
whereas more massive stars instantaneously end their lifes collapsing directly into a stellar black hole. 
For the sake of simplicity, we assume that each MS deposits 
both energy and mass into the ISM in two different phases,
i. e. at their birth and at the end of their lives (exploding as type II SNe), in equal proportions. 
Each MS of initial mass $m_{\rm ini}$ is assumed to deposit through its life an amount of mass
equal to $\eta~m_{\rm ini}$ with $\eta=0.9$ (therefore an amount $0.45~m_{\rm ini}$ is deposited immediately after its birth in the host cell.
This implies that each of these sources ends its life leaving a remnant with mass equal to 10\% of its initial mass. 
The feedback deposited in the pre-SN phase is aimed at accounting for the amount deposited 
in the stellar wind phase, during which MS are known to represent non-negligible sources of heating (see \citealt{ros14,cal15}). 
The pre-SN feedback channel is a fundamental one, as without including it, the first form of heating will be SN explosions,
which do not occur until after a few Myr, i. e. in our case,  at the end of the lifetime of $40 ~\rm M_{\odot}$ stars.
Due to the high gas densities reached in our simulation, this time is long enough to achieve an overly fast increase of the stellar mass.
Moreover, the role of pre-SN feedback in stellar complexes turned out to be more important than SNe, in particular in dense stellar aggregates (\citealt{hop10}). 
We assume that the energy deposited by each MS thoughout its entire life is proportional to its initial mass. 
Therefore, each MS deposits a total amount of thermal energy expressed as
\begin{equation}
E_{\rm th} = \frac{\eta~m_{\rm ini}}{10~\rm M_{\odot}} \cdot 10^{51} \rm erg. 
\end{equation}
For what concerns the pre-SN phase, a deposited energy increasing with mass is supported by early stellar
wind models in which the total mechanical power of the wind 
\begin{equation}
E_{wind} = \int L_w dt = \int \frac{1}{2}\dot{M} v_w^2 dt,
\label{eq_wind}
\end{equation}
(e. g., \citealt{wea77}, where $v_w \sim 1000$ km s$^{-1}$ is the typical wind velocity and the integral of Eq.~\ref{eq_wind} is performed over the period of the pre-SN phase) is mostly in the form of thermal energy. 

On the other hand, the scaling between SN energy and initial mass is currently unknown. 
\cite{suk16} proposes an interesting model that physically accounts for the explosion of type
II SN across a wide range of mass (between 9 $\rm M_{\odot}$ and 120 $\rm M_{\odot}$).
The models are single-metallicity, one-dimensional, they do not include any 
effects of rotation and are calibrated on two known SN remnants. As stated in their paper, the model is subject to various
uncertainties (concerning mostly mass loss and nucleosynthesis). The dependence of their results on key quantities, such as
mass loss rates and explosion parameters and dimensionality,  is still uncertain \citep{wan18}.

Due to its fluctuating behaviour, a relation between SN energy and mass similar to the one of \cite{suk16} is perhaps expected to produce 
more limited effects of SN feedback with respect to our assumption. This topic needs to be investigated in the future.  

A delayed cooling scheme is used to prevent overcooling for both pre-SN and SN feedback, as described later in Sect.~\ref{sec_del}. 
The prescriptions adopted for pre-SN ejecta from MS are summarised in the top-right panel of Fig.~\ref{fig_fee}. 
The quantity shown by the dashed line is the difference between the initial stellar mass $m_{\rm ini}$ and the mass released as pre-SN feedback, i. e.
$0.55~m_{\rm ini}$ as adopted in our simulations and compared to the analogous quantity obtained in pre-SN models of stars of various masses by \cite{ren17}, 
expressed as the stellar mass at the end of the pre-SN phase. 
As visible from this panel, our prescriptions for pre-SN mass ejection are in excellent agreement with the results of \cite{ren17}. 

Our choice to split in half the energy released by each MS is supported by a few studies 
that have shown that the energy delivered 
by massive stars in the pre-SN phase is comparable to that released by SN explosions (\citealt{cas75, ros14, fie16}). 
The only difference between 
our pre-SN and SN feedback is therefore only the timing, since for each MS, half of both its energy and mass
deposition occurs at its birth and half at its death. 
It is important to stress that our model represents a simplified 
feedback scheme designed to account for the effects of pre-SN and SN feedback.  
The effect of this instantaneous release of energy at the birth is presumably different
from the one of a more realistic feedback model, where the deposit of energy is spread out over the star's lifetime.
Moreover, the response of the gas is likely to be different from the one of a more complete pre-SN
modelling, that includes effects such as stellar winds, photo-heating and injection of momentum.
Our results indicate that at our resolution, our feedback implementation is efficient in regulating star-formation, 
but in the future, it needs to be improved in various aspects, which include a more gradual release of energy and other of the processes mentioned above. 

As for the amount of metals released by each massive star, we calculate the total metal yield $y_{\rm Z}$,
defined as the ratio between the amount of ejected metals and the initial mass,  
for stars of various masses and metallicities by means of the tables of \cite{woo95}. 
\cite{woo95} calculated nucleosynthetic yields from SN models for a large amount of isotopes, for an extended grid of stellar masses, including stars from 11 $\rm M_{\odot}$ to 40 $\rm M_{\odot}$,
and metallicities from $Z=0$ to solar. 
The calculated yields are shown as functions of initial mass and metallicity in the bottom-right panel of Fig. ~\ref{fig_fee}. 
The amount of metals ejected by each single star is expressed 
by a polynomial fit to the relation obtained by \cite{woo95} at $Z>10^{-4}~Z_{\odot}$, expressed as 
\begin{equation}
y_{\rm Z} = \Sigma_{k=1}^6 f_k ~ (m_{\rm ini}/\rm M_{\odot})^k
\label{eq_yield}
\end{equation}
with $f_k = [0.0108, -0.0026, 0.00026, -8.80870~10^{-6}, 1.22852~10^{-7},$ and $-0.5791]$. 
This fit is in excellent agreement with the expression obtained by other authors, considering other sets of yields (see \citealt{ino12}). 

The yields computed by \cite{woo95} at Z=0 are different from most of the sets at larger metallicity. 
However, our use of the interpolation in Eq.~\ref{eq_yield} should not represent a major concern. We have verified this by
calculating the fractional difference between the integrated yields per stellar generation (defined as in Eq.
1 of \citealt{vin16}), performing the integral in the massive star regime ($8 \rm M_{\odot} \le m \le 40 \rm M_{\odot}$) at
Z = 0 and at Z = 0.0001 $Z_{\odot}$. The fractional difference between the two integrated yields
amounts to  0.36, therefore our overestimate of the yields in a zero-metallicity stellar population
is well within a factor 2. 

\subsubsection{Delayed cooling for feedback of single stars}
\label{sec_del}
In simulations, in high density regions of the ISM such as the typical cold, star forming gas, 
the energy returned by MS can be artificially radiated away extremely quickly \citep{kat92} due to efficient cooling. 
The primary consequence of this phenomemon is an inefficiency of stellar feedback and an unregulated 
growth of stellar mass, resulting in strong disagreement with the most fundamental properties of galaxies and stellar systems.
This effect is due to a number of reasons, including inadequate resolution and a variety of  physical processes associated with stellar feedback
such as radiative heating from young stars \citep{gee16}, radiation pressure \citep{mur10} and other non-thermal processes such as
magnetic fields, turbulence and cosmic rays (see \citealt{tey13}; \citealt{far22} ). \\ 
Various methods have been proposed to prevent immediate radiative loss of the energy injected by MS (e. g., \citealt{tha00}, \citealt{age13}, \citealt{ros17}).
One way to prevent overcooling is by increasing resolution, where possible.  
\cite{kim15} modelled the evolution of a SN remnant to find the required optimal cell size to prevent overcooling. They found that in a uniform medium, the ideal
resolution has to be 3 times smaller than the quantity $r_{\rm SF}=22.1~n_0^{-0.4}$,
where $n_0$ is the gas number density in units of cm$^{-3}$. 
In our case, the maximum density can be of the order of $n_{0}\sim 10^5$ cm$^{-3}$; fully resolving the SN remnant at
such high density requires a resolution attainable only at extremely high redshift, well before the beginning of star formation. 
Besides increasing resolution, other sub-grid methods are used to prevent local overcooling. 
In this work, we temporarily switch off cooling in suitable cells, starting from the general method proposed by \cite{tey13}, 
in which the feedback is released as thermal energy and simultaneously stored in a tracer, passively advected 
passively advected with the flow. 
Each time a MS is born or a SN explodes, besides injecting the thermal energy of the star into the host cell,
a 'non-thermal' energy tracer is also accumulated on the grid in the form of a passive scalar $\rho_{\rm NT}$, 
associated with an unresolved turbulent energy. 
In the native implementation, 
created mostly for modelling star particles, 
radiative cooling is switched off in each cell in which the local non-thermal velocity dispersion  $\sigma_{\rm NT}$ is
above a certain limit $\sigma_{\rm min}$. 
Moreover, $\rho_{\rm NT}$ is assumed to decay on a dissipation timescale $t_{\rm diss}$. 
In this formalism, the adjustable parameters are 2, i. e. $t_{\rm diss}$ and  $\sigma_{\rm min}$. 
As for the dissipation timescale, the original choice of \cite{tey13} was to set $t_{\rm diss}=10$ Myr, i.e. of the order of the typical molecular cloud lifetime.
In our simulations we choose the resolution-dependent approach proposed by \cite{dub15} for computing  $t_{\rm diss}$:
\begin{eqnarray}
t_{\rm diss}&=& 0.82 \left( \eta_{\rm SN} \over 0.1\right)^{-1/3} \times \left( \epsilon_* \over 0.01\right)^{-1/3} \times \left( N_{\rm cell} \Delta x \over 4\times 10 \, {\rm pc} \right)^{2/3}\nonumber\\
	&\times & \left( n_0 \over 200 \, {\rm cm^{-3}}\right)^{-1/6} \rm \, Myr \, .
\label{eq_diss}
\end{eqnarray}

It is worth noting that in the original formula for $t_{\rm diss}$ of \cite{dub15}, $\eta_{\rm SN}$ represents 
the mass fraction of stars ending up into a type II supernova and depends on the adopted stellar IMF. In the case of a \cite{sal55} IMF,  
a value of $\eta_{\rm SN}=0.1$ 
corresponds to $10^{51}$ ergs of SN energy released for every 100 M$_{\odot}$ of stars formed in the simulation. 
In the case of a \cite{kro01} IMF, the appropriate value for this quantity is $\eta_{\rm SN}=0.15$. 

Our model is in a very different regime (in terms of stellar mass resolution) from the one of \cite{dub15}; 
without any knowledge of the appropriate value for this quantity, 
in our implementation we adopt an empirical model and use a value of 0.9. 
In practice, this choice corresponds to a factor $\sim 0.6$ in the value of $t_{\rm diss}$.  
Therefore, in our case, assuming $\eta_{\rm SN}=0.9$, $\epsilon_* =0.1$, $N_{\rm cell}$= 4,  $n_0 =10^5$ cm$^{-3}$ and
for a maximim resolution $\Delta x=0.2$ pc (obtained at $z=10$), 
we have $t_{\rm diss} \sim 0.005$ Myr.

In the native implementation, 
the value adopted for $\sigma_{\rm min}$ is 100 km s$^{-1}$, 
corresponding to 0.1\% of the  specific energy injected by a single SN \citep{ros17} and implying that
the cells in which the cooling is reinstated after the delay are all the ones presenting $\rho_{\rm NT}<0.001 \rho$. 
From a physical point of view, this is equivalent to shutting down the cooling in cells in which the non-thermal velocity dispersion is still above
a certain threshold  (\citealt{tey13,dub15,ros17}).

It is important to note that in our formalism, we use  $\rho_{\rm NT}$ as the proxy of a turbulent energy density only because
we are assuming a uniform specific energy injected by MS.
For this reason, a mass tracer is equivalent to an energy tracer.
In this context, it is therefore legitimate to track the dissipation of the non-thermal energy through the passive scalar
$\rho_{\rm NT}$. 

In lower-resolution simulations or in the case of a different modelling of stellar feedback (such as
a gradual release of energy of single stars, e. g. in the pre-SN phase), 
it may be frequent to have a few MSs ejecting mass and energy in a single cell in which the density is very high, therefore 
$\rho_{\rm NT} \ll 0.001 \rho$.  
One possibility to overcome this is to elaborate a fine-tuning of the parameter $\sigma_{\rm min}$, which is however time-consuming, given 
the expected order-of-magnitude variations in the optimal choice as a function of cell size and feedback implementation and considering
that such choice might depend on resolution in a non-trivial manner. 
We propose instead a simpler approach to prevent overcooling and assume that, in cells where single MS have deposited energy and mass,
$\rho_{\rm NT}$ is posed equal to $\rho$, to switch off cooling for a limited amount of time. 

By construction, this approach is versatile as it is equivalent to inizializing the flag used for switching off radiative cooling $f_{\rm NT}=\rho_{\rm NT}/\rho$ to the maximum value of 1, independent on the amount of ejecta present in the cell. 
By means of a few low-resolution runs, 
we have carefully verified that this new approach ensures 
that the stellar feedback is correctly accounted for and that the stellar mass of our system is satisfactorily reproduced.
On empirical grounds, the choices described in this section enable an effective model which allows us to 
achieve the desired results, i. e. an efficient feedback at our resolution and considering our ingredients.

\subsubsection{Asymptotic Giant Branch stars}
Even if AGB do not contribute substantially to the heating of the ISM, in the case of a standard IMF such as the one adopted here,
their overall contribution in terms of ejected mass is dominant with respect to other stellar sources. 
In our model, individual AGB are assumed to release all their mass at the end of their lifetimes. To compute the total mass
restored by an AGB, we consider an analytic, linear fit to the final-to-initial mass relation (FIMR) of \cite{cum18}, based on
an analysis of an observational sample of white dwarfs from various star clusters. The study of  \cite{cum18} was based on the use of 
two different sets of white dwarf progenitor masses and indicates  a weak dependence of the results on the used model.
The results of their study are shown in the bottom-left panel 
of Fig.~\ref{fig_fee} as black diamonds, reported with their error bars on the derived final mass values. The analytic fit to the FIMR used in our work is 
\begin{equation}
m_{\rm fin} = 0.56 + 0.084~m_{\rm ini}, 
\label{eq_mfin}
\end{equation}
shown in the same panel by the thick dot-dashed dark cyan line. 
The fit is accurate in the range $4 \le m/\rm M_{\odot} < 8$, whereas it leads to a slight 
overestimate of the final mass for stars of 3 $\rm M_{\odot}$ \footnote{In their study, \cite{cum18} find a deviation from a linear relation at initial masses $< 3~\rm M_{\odot}$, i. e. in a mass range not considered here for stellar feedback.} 
The relation of eq.~\ref{eq_mfin} is used to derive the mass ejected by each AGB star, 
computed as the difference between the initial and final mass, $m_{\rm ini}-m_{\rm fin}$. 

We neglect the contribution of AGB to the heating of the ISM and assume that the amount of metals
they restore is $Z_{\rm ini} \cdot (m_{\rm ini}-m_{\rm fin})$, where $Z_{\rm ini}$ represents the metallicity of the ISM at their birth, stored
into a variable at the birth of each individual star, along with the birth time. 

In the present study, we do not consider the feedback from type Ia supernovae. For a \cite{kro01} IMF, as adopted
here, the total number of type Ia SNe per stellar mass in a simple stellar population across a Hubble time is $\sim 10^{-3}$
(\citealt{lac21}), therefore significantly lower than the number of MS ($\sim$0.01 per stellar mass), which will dominate
the global budget of energy and metals restored into the ISM.

\begin{figure*}
	\includegraphics[width=18.cm,height=7.cm]{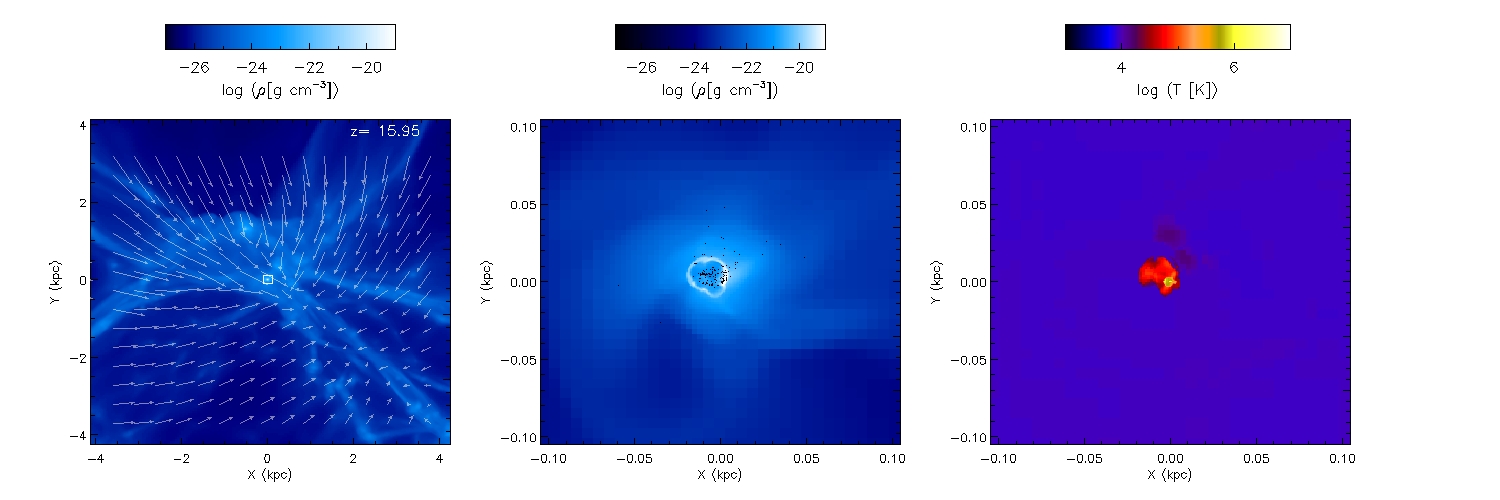}
\caption{Left: density map in the x-y plane of a limited region of transverse physical size $\sim 8$ kpc 
containing the first stars at $z=15.95$. 
The map is centered on the region of maximum temperature of the entire volume,
which is also one of the highest density locations. The arrows represent the velocity field of the gas.
Middle panel: zoomed slice density map  of the region enclosed in the white square shown in the left panel.
The black dots show the positions of the stars and of the stellar particles. Left map: slice temperature
map of the same region as shown in the middle panel. All maps are computed as mass-weighted averages along the line of sight.} 
\label{fig_0018}
\end{figure*}
\begin{figure*}
	\includegraphics[width=18.cm,height=7.cm]{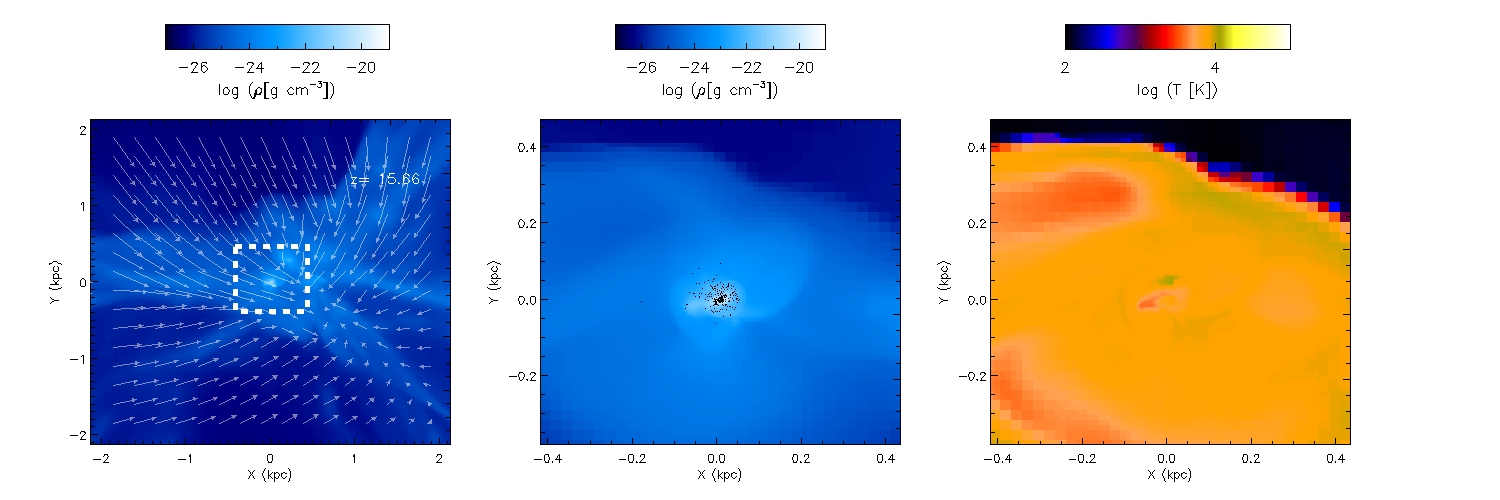}
\caption{Left: density map in the x-y plane of a limited region of transverse physical size $\sim 5$ kpc 
that contains the first stars formed in our high-resolution simulation, computed at $z=15.66$. The middle panel shows a
zoom on the region enclosed by the white dashed line of the left panel. The black dots are as in Fig.~\ref{fig_0018}. 
The right panel shows a slice temperature map of the same region as in the middle panel. 
All maps are computed as mass-weighted averages along the line of sight.} 
    \label{fig_0019}
\end{figure*}

\section{Results}
In this section, we present an overview of main results from our simulations. 
First, we show results concerning the global evolution of the simulated
system, focusing in particular on the gas and stellar components.
We also present the physical properties of the stellar
clumps in place at the final redshift of $z=6.14$.
More detailed analysis will be presented in follow-up works.

\subsection{Global evolution of the system}

\subsubsection{Evolution of the gas}
 In our simulation, star formation begins at $z=15.95$, corresponding to a cosmic time of 0.251 Gyr. 
Fig. ~\ref{fig_0018} shows slice density and temperature maps, calculated on the x-y plane, of a limited region of
the computational volume at $z=15.95$ that includes the site where star formation ignites for the first time. 
At this redshift, the maximum spatial resolution is of $\sim 0.2$ pc. 
The slice map reported in the left panel of Fig. \ref{fig_0018} represents a region of physical size $\sim 8$ kpc 
and outlines the filamentary nature of the density structure. 
\begin{figure*}
	\includegraphics[width=16.3cm,height=22.cm]{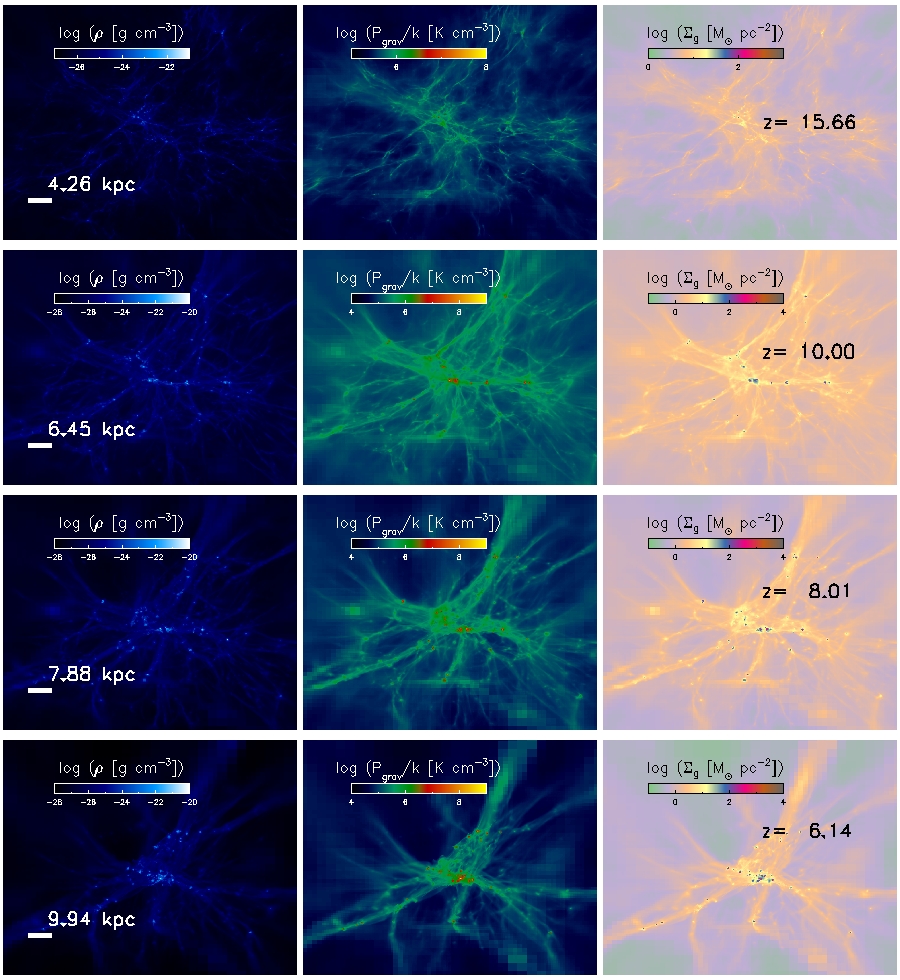}
\caption{Global evolution of the main properties of the gas component in the highest-resolution simulation. 
The maps describe the density (left column), gravitational pressure (middle column) and gas surface density (right column)
distributions in the central, zoomed-in region of the box,
computed at four different redshifts (from top-to bottom: $z=15.66, z=10, z=8.01, z=6.14$).
The horizontal white solid lines shown in the left column-panels indicate the physical scale.
The density maps are computed as mass-weighted averages along the line of sight.
The other cases are column density maps, in which the quantities are summed along the line of sight.}
        \label{fig_gas_multi}
\end{figure*}
The map is centered on the region of maximum temperature of the entire volume,
which is also one of the highest density locations. The density ranges between $10^{-26}$ g cm$^{-3}$ and 
$10^{-19}$ g cm$^{-3}$; the latter value corresponds to maximum density of $\sim 10^{5}$ particles cm$^{-3}$.
The velocity field follows the extension and shape of the dense filaments, characterised by
$\rho> 10^{-23}$ g cm$^{-3}$, and shows that the overall motion of the cold gas is directed towards the central region of the plot. 
A few dense knots are visible along some of the filaments and the maximum density is reached in one in particular,
visible at the centre of the left panel and indicated by an open square. The central panel of Fig. \ref{fig_0018}
is a zoomed density map of this  $\sim~0.2$ kpc-wide 
region. A diffuse distribution of stars and stellar particles is visible, shown as black dots and particularly dense at the centre,
whose features will be described in more detail later. The central stellar component is embedded in 
a $\sim$30 pc-wide cavity, surrounded by a dense (with density as high as $10^{-19}$ g cm$^{-3}$), thin shell of thickness of a few parsec.
This shell surrounds a hot expanding bubble powered by the feedback of the newly born MS. 
The effects of the feedback are shown also in the temperature slice map, shown in the right panel of Fig. \ref{fig_0018}.
A $\sim10$ pc-wide, very hot region with temperature $>10^6$ K is visibile at the centre, surrounded by a more extended, confined structure
with T $\sim 10^5$ K, whose shape is reminescent of the cavity surrounded by the thin shell seen in the central panel.
The presence of such a hot, rarefied (with respect to 
the shell) cavity confirms that our pre-SN feedback model is efficient at this resolution. 

Fig. ~\ref{fig_0019} shows slice density and temperature maps of the same region computed a few Myr later, at $z=15.66$, corresponding to a cosmic time of 0.258 Gyr.
The region shown in the left panel is centered on the highest density cell of the computational volume.
At the centre, there is also the same stellar component present in Fig.~\ref{fig_0018}, but slighly grown in mass and number of stars.
The central panel is a zoomed density map of the 1 kpc-wide region enclosed within the dashed-line square shown on the left.  
The cavity visible at the previous time has disappeared and the gas has recollapsed into a central clump, in which most of the stars are 
embedded. The central density, of the order of $10^5$ cm$^{-3}$, is comparable to the densest regions in local molecular clouds (e. g., \citealt{dob14}).
The temperature map on the right shows that the gas has cooled down to lower temperatures than shown in Fig.~\ref{fig_0018},
in this case of the order of T $\sim 10^4$ K in the central regions.

Fig.~\ref{fig_gas_multi} shows the global evolution of the main properties of the gas component
in the highest-resolution simulation.
The maps describe the density, pressure structure and surface density distribution of the gas in the central, zoomed-in region of the box,
computed at four different redshifts.

In the middle panels, we report the gravitational pressure of the gas,
expressed as the gravitational force per unit area, i.e. $P_{\rm grav} = G \times \Sigma_{\rm gas}^2$,
where $G$ is the gravitational constant and $\Sigma_{\rm gas}$ is the gas surface density.
In virialised systems like our clumps, the gravitational pressure equals the kinematic pressure $\rho~\sigma^2$
(where $\sigma$ is the velocity dispersion), and is suited to infer the amount of turbulence present in the system 
(\citealt{elm97}, \citealt{ma20}). 

The three maps in the top row describe the properties at $z=15.66$, i. e. soon after the onset of star formation.
At this epoch, the seeds of a network of dense, low-pressure, cold (with typical temperatures of $\sim 10^3$ K)  filaments
with surface densities $\sim 10 ~\rm M_{\odot}/pc^2$ are present.  
At a later time ($z=10$, second-from-top row of Fig.~\ref{fig_gas_multi}) 
 a handful of compact, dense ($\rho > 10^{-22}$ g cm$^{-3}$) clumps are visible at the centres of the dense gas filaments. 
The clumps have $\Sigma_{\rm gas} \sim 10^2 \rm M_{\odot}/pc^2$, size of $\sim 0.1$ kpc and are characterised by high pressures,
P$_{\rm grav}$/k  $\sim 10^6$ K cm$^{-3}$, 
comparable to turbulent star-forming regions in local molecular clouds (e. g., equivalent to
a cloud with a particle density of $100$ cm$^{-3}$ and a turbulent velocity of $10$ km/s) and several orders of magnitudes
larger than the warm ISM in the Milky Way disc, characterised by P$_{\rm grav}$/k  $\sim 10^3$ K cm$^{-3}$.  
At later epochs, the overall properties of the filaments do not evolve much but the number of collapsed gas clumps
increases considerably. 
At the final redshift of the simulation, several tens of clumps are present in the zoomed-in region, with increased
maximum surface densities, now in excess of $10^3 \rm M_{\odot}/pc^2$. Also the pressure of the most central clumps
has increased, to reach P$_{\rm grav}$/k  = $10^8$ K cm$^{-3}$ at $z=8$. 
Locally, these extreme pressure values correspond to the ones of the most highly-pressurised medium in 
molecular clouds in the turbulent centres of local star-forming galaxies \citep{sun18,che20} or in local starbursts and mergers
(\citealt{mol20} and references therein), generally higher by a few orders of magnitude than the Milky Way. 
Such regions are rare locally, but are expected to be more frequent at high redshift (\citealt{kru14,ma20}). 

In conclusion, the large-scale properties of the gas component are qualitatively not too different from 
the results of other cosmological zoom-in simulations performed at various scales (see, e. g., \citealt{dub15}; \citealt{sch15}; \citealt{zhu20};
\citealt{cos18} and others; for a review, see \citealt{vog20}). 
The most remarkable aspect of our simulations concerns the high spatial resolution, 
which allows us to resolve the cold gas up to densities and pressure values  
generally not achievable in simulations with typical resolution of 10 pc or more. 

\begin{figure*}
	\includegraphics[width=18.cm,height=18.cm]{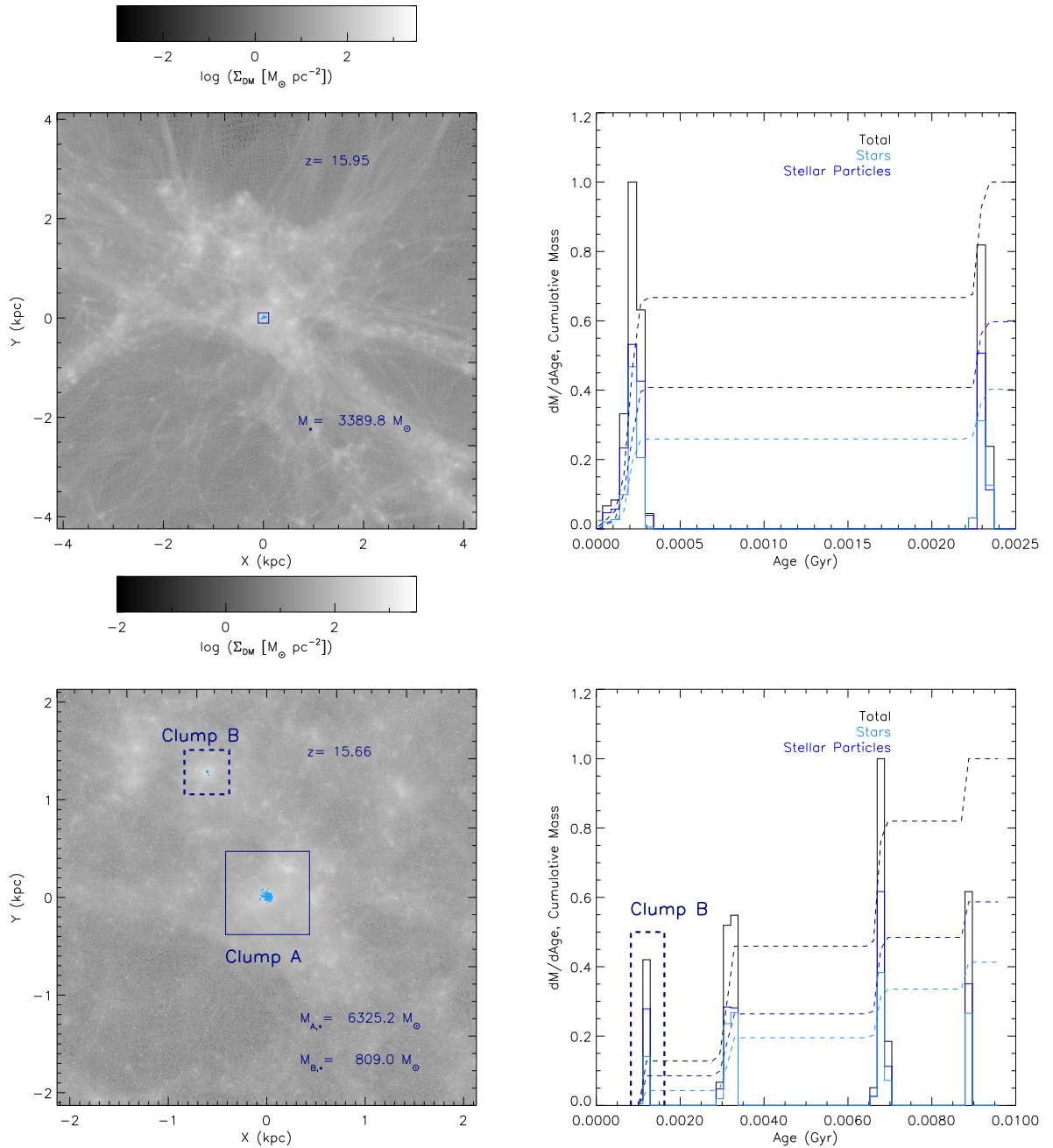}
\caption{Properties of the first stars formed in our highest-resolution simulation.  
The top-left panel shows a projected density map (where the mass has been summed along the line of sight) 
of the DM distribution of transversal size of $\sim 8$ kpc at $z=15.95$, where 
the stars are indicated by the cyan dots at the centre of the map. 
The top-right panel shows the differential (solid lines) and cumulative (dashed lines)  
mass distributions of age computed for the individual stars (cyan), stellar particles (blue) and the sum of the two (black).
The bottom-left panel shows a projected density map of the DM distribution of transversal size of $\sim 4$ kpc at $z=15.66$, where 
the stars are indicated by the cyan dots. The oldest, main stellar clump A is enclosed by the blue solid box,
whereas the youngest (Clump B) is enclosed by the blue dashed box. The stellar masses of the clumps are reported on the righ-bottom.
The bottom-right panel is as the top-right one. The distributions for Clump B are enclosed in the blue dashed box.}
\label{fig_age_stars}
\end{figure*}

\subsubsection{Evolution of the stellar component}

The top panels of Fig. ~\ref{fig_age_stars} illustrate some properties of 
the first stars, formed in our simulation at $z=15.95$. 
The top-left panel shows the location of the stars with respect to the DM distribution, originating at the centre of an overdensity
with DM surface density 
$\Sigma_{\rm DM}\sim 10^3~ \rm M_{\odot}$ pc$^{-2}$ and enclosed by the 0.2 kpc-wide blue open square. 

The stellar age distribution at this redshift is shown in the top-right panel. 
Two populations have originated, with a difference in age of $\sim$2 Myr and with a total stellar mass of $\sim~3.4 \times  10^3~\rm M_{\odot}$.
The age distribution shows both the normalised differential and cumulative contribution to the stellar mass of individual stars
and stellar particles. Most of the mass is in the form of stellar particles, which reflects our assumption of storing
all the stars formed in the lowest stellar mass bin (ranging from $0.1~\rm M_{\odot}$ to $3~\rm M_{\odot}$) into them.
For the \cite{kro01} stellar IMF chosen in this work, the normalized mass fraction in this bin is $\sim$ 0.6, corresponding to 
the final value of the cumulative age distribution of stellar particles. This value occurs at 0.0023 Gyr, which is the age
of the oldest stars present at this epoch.  

The bottom panels of Fig. ~\ref{fig_age_stars} show the properties of the first-born stellar populations a few Myr later, at $z=15.66$.
Two distinct stellar aggregates are visible in the DM density map on the bottom-left. 
The older, more massive population lying at the centre and
a younger component are highlighted by the solid and thick dashed blue box, respectively, located at a relative distance of
$\sim 1.5$ kpc. 
The total stellar mass is $6400~\rm M_{\odot}$, 13\% of which is in the lower-mass, younger clump (clump B in Fig.~\ref{fig_age_stars}).  

The stellar age distribution is shown on the bottom-right of Fig. ~\ref{fig_age_stars}.
At this time, 4 main populations are present, the 3 oldest of which are in Clump A. 
All the stellar populations are younger than 9 Myr. 
The most massive, still living stars have mass of 25 $\rm M_{\odot}$ and, considering our adopted stellar lifetimes, 
will die exploding as SNe (and restoring the first metals) $\sim$ 9 Myr after their birth.
At this epoch, no massive star has died yet, therefore no metals have been
returned, Since metals are returned only with SN explosions.  
This explains why the gas at this epoch is still metal-free, as we will see later, and shows how the star formation is 
regulated by pre-SN feedback only. 

\begin{figure*}
        \includegraphics[width=16.3cm,height=22.cm]{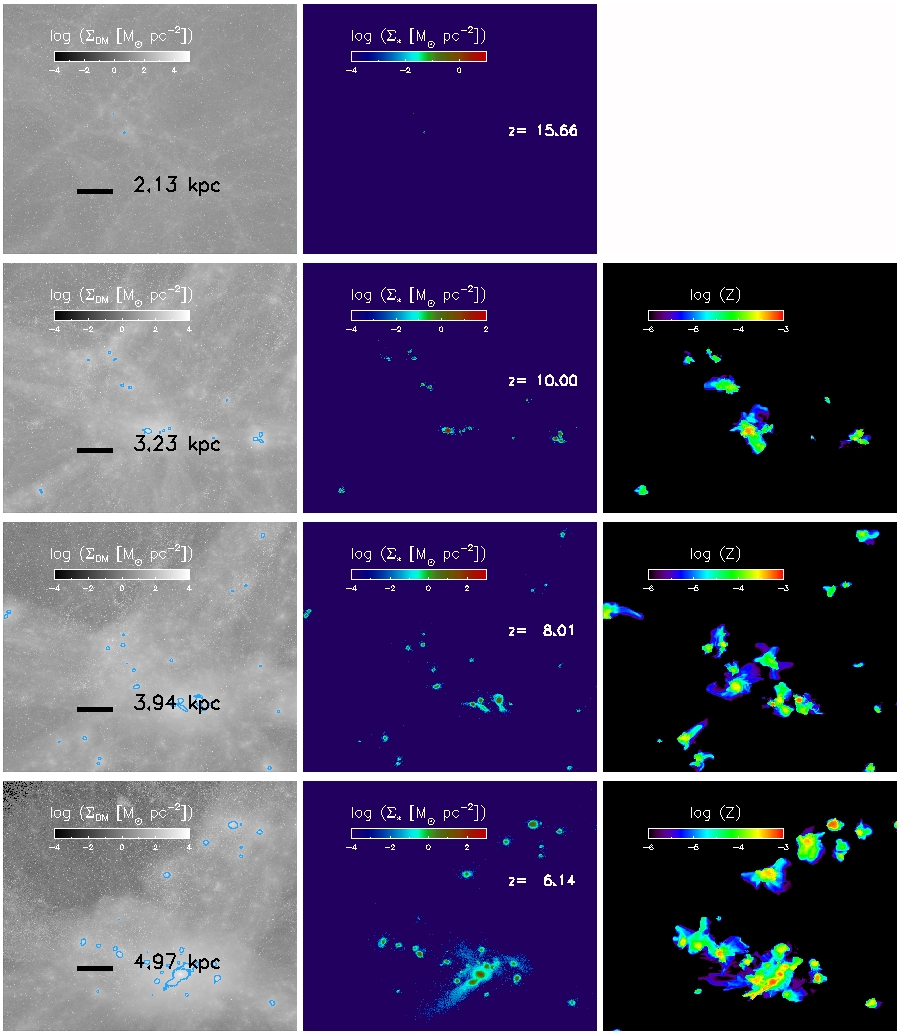}
\caption{Global evolution of some of the main properties of the stars in the highest-resolution simulation. 
Projected mass density maps of the dark matter and of the stars at various redshifts are shown in the left and middle column,
repspectively. 
The cyan contours in the left column indicate regions where the stellar surface density is $>0.1~\rm M_{\odot}$pc$^{-2}$.
On the right, slice gas metallicity maps are reported at three different redshift values. 
No metallicity map is shown at $z=15.66$ (top row) since at this redshift the gas is sill metal-free (see the text for further details).
Each DM and stellar surface density map is calculated by means of a sum of the stellar and DM mass along the line of sight, respectively.
The metallicity maps  are computed as mass-weighted averages along the line of sight.}
    \label{fig_maps_stars}
\end{figure*}

\begin{figure*}
	\includegraphics[width=12.cm,height=12.5cm]{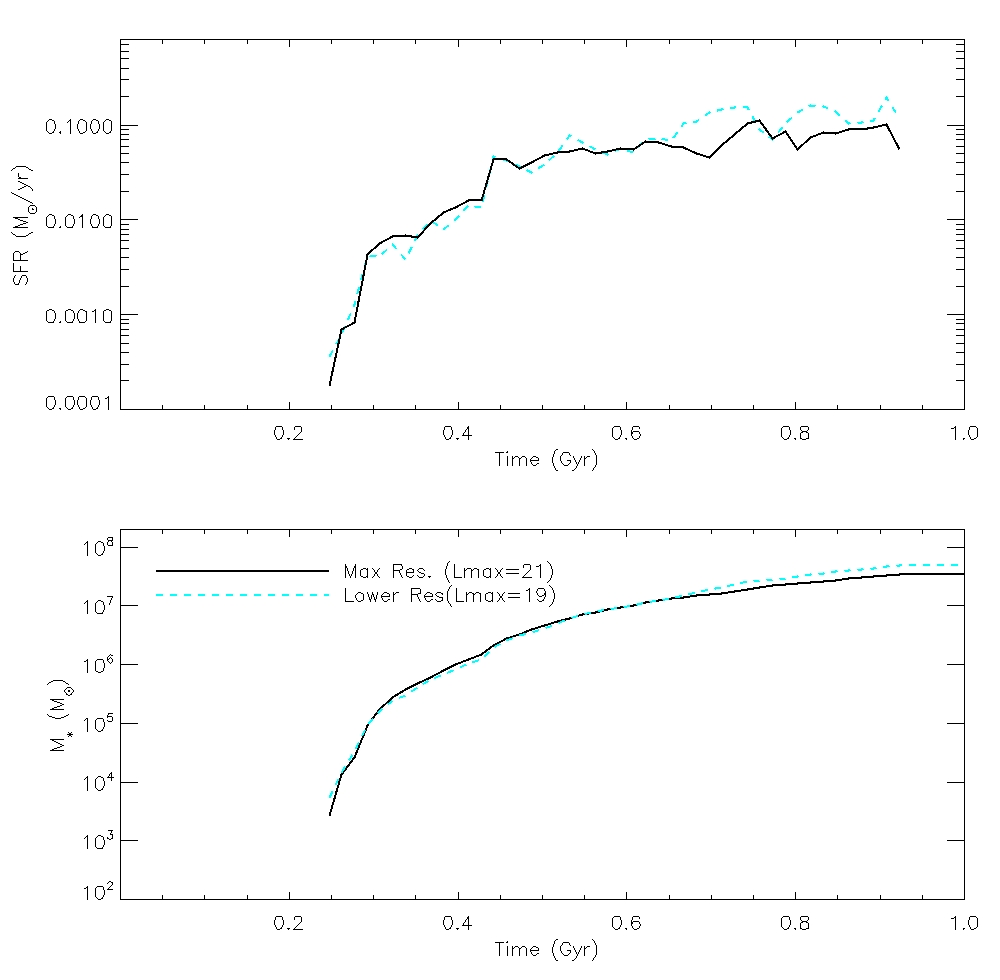}
    \caption{The upper and lower panels show the evolution of the total star formation history and of the cumulative stellar mass
    in our high-resolution (solid black line)
    and lower resolution (cyan dashed line) simulations.}
    \label{SFH}
\end{figure*}

Fig.~\ref{fig_maps_stars} shows the evolution of some of the main properties of the stellar component
in our highest resolution simulation. 
The maps describe the DM surface density (left column), stellar density (central column) and 
gas metallicity distribution (right column) in the central, zoomed-in region of the box,
computed at the same redshifts as in Fig.~\ref{fig_gas_multi}.

As cosmic time evolves, more and more stellar clumps form at the centre of the densest DM halos   
with mass density $>10^2~\rm M_{\odot}$ pc$^{-2}$. At the same time, an increasing fraction of the
volume is enriched with metals due to exploding SNe. The maximum metallicity is $Z=10^{-3}$, reached in regions
where the largest stellar clumps are located. A more detailed discussion of the properties of the stellar
clumps will be performed later, in Sect.~\ref{sec_clumps}.

Fig.~\ref{SFH} shows the star formation history (upper panel) and cumulative total stellar mass (lower panel)
of our highest resolution simulation, compared with the results of an analogue run with the same setup but lower resolution.
In the latter run, the maximum refinement level is lmax=19, corresponding to a factor 4 wider cells at the highest resolution.
This figure is useful also to check the numerical converge of our runs. 
In the two simulations, SF starts at a remarkably similar epoch and, after a continuous increase until a cosmic time of
0.5 Gyr, the SFR values settle to comparable values, of the order of $\sim 0.1 ~\rm M_{\odot}$ yr$^{-1}$.
Likewise, the evolution of cumulative stellar mass in the two runs is remarkably similar. At the final
time (0.92 Gyr), the total final stellar masses are consistent to within a factor 1.4, with a larger value
in the lower resolution run. Considering the intrinsic stochasticity of SF and the resolution-dependent
implementation of some of our prescriptions, such as the dissipation timescale for delayed cooling,
the agreement between the results of the two runs is satisfactory.

%

%
\begin{figure*}
         \includegraphics[width=8.9cm,height=21.5cm]{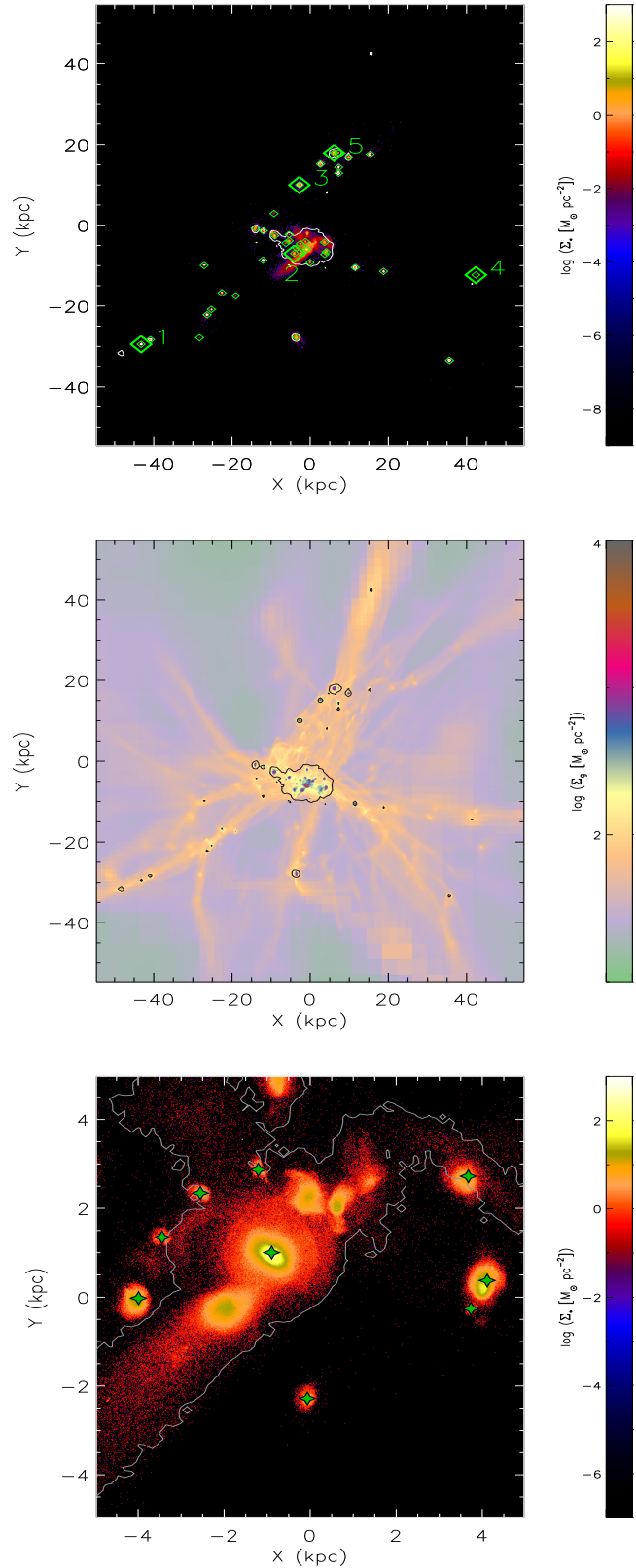}
\caption{Upper panel: projected stellar density map at redshift $z=6.14$ of the central region of the box, with transversal physical size of
$\sim 100$ kpc. 
The small green open diamonds mark the positions of the stellar clumps identified with fof (see the text for
further details). 
The positions represent the mass-weighted averages of the x- and y- coordinates of all the individual stars and stellar particles
belonging to each stellar clump. 
The big green diamonds mark the positions of a few representative stellar clumps shown in Fig.~\ref{fig_clump1}, \ref{fig_clump4},
\ref{fig_clump2}, \ref{fig_clump3}, and \ref{fig_clump5}. 
The white contours enclose the regions with DM surface density $>10^{2.5}~\rm M_{\odot}$pc$^{-2}$.
Central panel: gas surface density map of the same region and with contour lines as in the upper panel.
Bottom panel: zoomed, $\sim$10 kpc-wide  stellar density map of the central region of the box. The green stars are on 
the centres of mass of the identified clumps. 
The grey contour encloses one group of stars identified by fof which includes a diffuse, central stellar
component, whose centre of mass coincides with the highest density point.
Each map and contour is calculated as a sum of the mass along the line of sight.}
    \label{fig_sig}
\end{figure*}

\subsection{Properties of the clumps}
\label{sec_clumps} 

In our simulations, the stellar clumps have been identified by means of the friends-of-friends algorithm 
fof\footnote{https://github.com/N-BodyShop/fof}, a versatile program created to find groups in N-body simulations.
The code is set to reject any group with less than a threshold number of members, defined by
the user. By definition, a group  identified by fof is one in which every particle 
 has at least one "friend" particle within a distance less than or equal to a "linking length".  
We made a few attempts to vary both the minimum particle number N$_{\rm min}$ and the linking length ll and
settled on those values for which the results were converging,  i. e., $\rm N_{\rm min}=64$ and ll=1 kpc,
respectively. 

The maps in fig. \ref{fig_sig} show the central, highest-resolution region containing
all the stellar clumps identified with fof at the end point of our simulation, i.e. at redshift $z=6.14$.
 

Some clumps show signs of distortions due to interactions with nearby structures, as well as  
extended tails, formed by stars lost during the interactions and 
along their orbital motions. The clumps with nearby companions are very often
connected to them via low-density 'bridges' of stars (bottom panel of Fig. \ref{fig_sig}). 

An elongated stellar component is visible at the centre of the stellar density map, including
a major fraction of the total stellar mass (1.8 $\times 10^7~\rm M_{\odot}$). This extended
distribution (Fig. \ref{fig_sig}, bottom panel) is characterised by density $\sim 10^{-1}~\rm M_{\odot}$ pc$^{-2}$ and is the result
of multiple mergers, interactions and fly-bys between clumps, which gave rise
to the stripping of a considerable amount of loosely-bound stars. The elongated stellar component extends
well beyond the extent of the dense gas. This outlines further its dynamical origin, i. e. that it originated
from stars shed by the clumps during their mutual interactions. 

The map in the middle of Fig. \ref{fig_sig} shows that each DM subhalo contains a dense gas component, generally located
at the centre of the halo. Most of the stellar clumps have a gas conterpart,
and the two components are broadly co-spatial. On the other hand, the DM sub-halos which do not show any
presence of stars are considerably rare.

The bottom panel shows a zoom on the central region of the box. 
The elongated  structure that includes the central clump is fragmented by fof in a few sub-clumps,
whose number depends on the adopted values for $\rm N_{\rm min}$ and ll. 
From a visual  inspection, it is clear that these sub-clumps are artifacts of the algorithm.
This incorrect pattern recognition is due the fact that here lies the most massive clump of the simulation (with mass $\sim 10^7~M_{\odot}$),
that it is interacting with other smaller systems and that the field is contaminated by extended stellar streams
produced by previous interactions. In our analysis, we have considered this extended central region as a single clump
and we have removed it from the fof clustering process.
With this precaution, fof  detects the same number of clusters among the remaining
stellar particles for any given choice of N and l above the mentioned threshold values.

Each stellar clump identified by fof lives within a DM halo. This suggests that 
each stellar clump is gravitationally bound; this issue will be addressed more specifically in a forthcoming work, extending the 
analysis performed with fof to a full phase-space analysis on the stellar particles (Pascale et al., in prep.). 

\begin{figure}
	\includegraphics[width=8.cm,height=11cm]{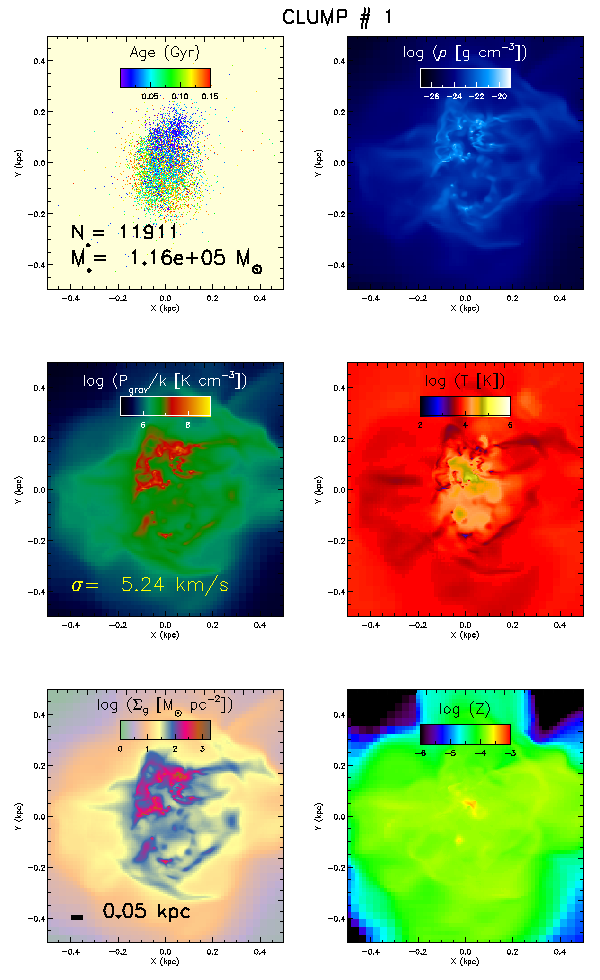}
\caption{Main properties of clump 1 of Fig. \ref{fig_sig} and slice maps of various properties of the gas, each one
centered on the centre of mass of the clump. Top-left panel: scatter plot of the stars,
colour-coded with their age. Top-right: gas density map. Middle-left: pressure map. Middle-right: temperature map.
Bottom-left: gas surface density. Bottom-right: metallicity map.
The density, temperature and metallicity maps are computed as mass-weighted averages along the line of sight.
For the gravitational pressure and surface density we show column density maps, in which the quantities are summed along the line of sight.}
    \label{fig_clump1}
\end{figure}

\begin{figure}
	\includegraphics[width=8.cm,height=11cm]{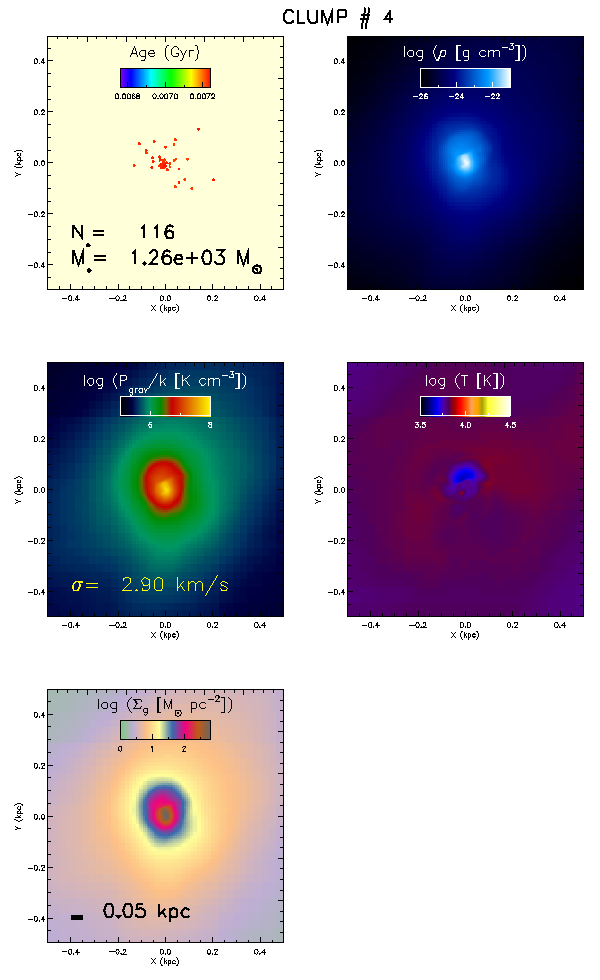}
\caption{Main properties of clump 4 of Fig. \ref{fig_sig} and slice maps of various properties of the gas, each one
centered on the centre of mass of the clump. Panels as in Fig. \ref{fig_clump1}.}
    \label{fig_clump4}
\end{figure}

Two representative cases of stellar clumps and of their gas counterparts are shown in Figures~\ref{fig_clump1} and \ref{fig_clump4},
showing various properties of the clumps (1 and 4 of the top panel of Fig. \ref{fig_sig}, respectively) and of the gas
surrounding each clumps. 
The stellar masses of clumps 1 and 4 are $1.16\times 10^5 \rm M_{\odot}$ and $1.26\times 10^3 \rm M_{\odot}$, respectively. 
Clump 1 has another object nearby with similar stellar mass. However, both clumps are in isolated regions, several
10 kpc away from the centre of the box, populated with a multitude of stellar systems.

Fig.~\ref{fig_clump1} shows that the youngest stars coincide with
the densest and highest-pressure (P$_{\rm grav}$/k  $>10^6$ K cm$^{-3}$) gas. 
The density and pressure maps show that the gas is turbulent. 
This is indicated also by the computed velocity dispersion of the cold gas. 
For each of the 5 clumps shown in Fig.~\ref{fig_clump1}, ~\ref{fig_clump2}, ~\ref{fig_clump3}, ~\ref{fig_clump4} and ~\ref{fig_clump5}, we have computed the 1D density-weighted velocity dispersion of the cold gas (with temperature $<200$ K) 
defined as: 
\begin{equation}
\sigma^2 = \frac{1}{3}\frac{\Sigma \rho_c [(v_x - \bar{v_x})^2 + (v_y - \bar{v_y})^2 + (v_z - \bar{v_z})^2]}{\Sigma \rho}
\end{equation}
(e. g., \citealt{she10}; \citealt{cal20}), where $\rho_c$, $v_x$, $v_y$ and $v_z$ are the density, 
the x-, y- and z-component of the velocity of the cold  gas 
in a cell respectively, 
whereas $\bar{v_x}$, $\bar{v_y}$ and $\bar{v_z}$ are the average x-, y- and z-component velocity values, respectively. 

The computed $\sigma$ is shown in the middle-left panel of Fig.~\ref{fig_clump1} and 
is in agreement with the velocity dispersion measured in local clouds, e. g. as traced by the relation between $\sigma$ and size observed in 
molecular clouds (\citealt{lar81}, \citealt{elm00}). 
Moreover, in most cases the obtained $\sigma$ values are larger from the expected dispersion due to thermal pressure only,
in our case of the order of $1$ km/s, corresponding to the sound speed at our assumed temperature floor ($T=100$ K). 
The highest-metallicity gas has $Z\sim 0.001$.\\
Clump 4 (Fig.~\ref{fig_clump4}) is composed of a considerably lower number of stars and the average gas density is significantly
lower than clump 1. The stars are much younger, the gas is much less perturbed as due to the lower
number of MS and 
it shows significantly lower maximum density, temperature and $\sigma$\footnote{Clump 4 presents a lower dynamic range in temperature with respect to the
other systems. In this case, $\sigma$ has been computed considering the gas with $T<5000$K.}. The gas is still metal-free. 

Other examples (clumps 2, 3 and 5 of of the top panel of Fig. \ref{fig_sig}) are shown in the Appendix.



\subsection{Comparison with D1+T1}
The total stellar mass of the simulated system at $z=6.14$ is 3.4$\times10^7~\rm M_{\odot}$. 
This value is in satisfactory agreement with the observed stellar mass of the D1+T1 system identified by \cite{van17a} which,
adding up the contribution of these two systems only, amounts to 2.3$\times10^7~\rm M_{\odot}$. 
Additional fainter star-forming knots at the same redshift have already been identified
and some of them have been spectroscopically confirmed \citep{van19}, with de-lensed magnitudes fainter than 32.
Currently, work is in progress to determine the physical sizes and stellar masses of
other 5 systems (Mestric et al., in prep.). 
Based on the brightness of the visually identified systems, it appears likely that the bulk of the stellar mass 
is dominated by the D1+T1 system. Moreover, no significant presence of dust is expected to 
be obscurating such low-mass systems, which might have major effects in the determination of basic quantities such as mass, age and SFR.

The clump sizes are computed as the half-mass radii of the surface density distributions, calculated assuming for
each system a circular shape.

The sizes of the simulated highest mass clumps are typically of the order of $\sim 100$ pc. 
On the other hand, the measured half-light radius of D1 is 44 pc. At face value, this implies that the average stellar density
of the simulated clumps is $\sim$1 order of magnitude lower than the observations.
Moreover, \cite{van19} report on a very dense star-forming region (with mass $~10^6~\rm M_{\odot}$ and radius
$<13$ pc) at the centre of D1, which fully qualifies as a young stellar cluster. 
In the current analysis, no such system is present in our simulation.  

The intrinsic SFR derived for D1 from the
SED-fitting at $3\sigma$ ranges between 0.4 $\rm M_{\odot}/$yr and $\sim78~\rm M_{\odot}/$yr, whereas
the best value is $15.8~\rm M_{\odot}/$yr \citep{van19}.
This value is the result of a SED-fitting analysis, showing degenerate solutions for
stellar mass, age, and SFRs \citep{van19}. 

Considering the tentative [CII] detection of D1 performed with ALMA, resulting in a [CII] luminosity
$L_{\rm CII} = 2.9~\times~10^6~L_{\odot}$ \citep{cal21}, such a high SFR causes D1 to deviate significantly from the tight 
SFR-$L_{\rm CII}$ relation observed in high-redshift systems \citep{car18}. 
The final SFR value of the simulated systems is lower than the best value reported
by \cite{van19} by more than two orders of magnitude. 
Were the SFR of D1 of the same order of magnitude as our predicted value,
this would make it much more consistent with other high-redshift systems in the observed SFR-$L_{\rm CII}$ relation
and alleviate the discrepancy reported by \cite{cal21}. 

In the simulations, a non-negligible fraction of the stellar mass is present in a low-density component,
which is the elongated structure visible in Fig. \ref{fig_sig}. It is expected that, due to its low surface brightness,
such a structure will be undetected in observations. 
As for the mutual distance between the stellar clumps, in our simulations
it can vary much from very low values in the case of tight interactions (right panel of Fig. \ref{fig_sig}) up to a few 10 kpc,
in case of the systems at the farthest distances from the centre. 
In the case of the real systems, estimating the physical distance at large separations is problematic
due to the strong variations of the magnification parameter $\mu$ throughout the observed field.
\cite{van17a} report on a source at the same redshift of D1 (as measured from the Lyman-$\alpha$ emission),
with an optical counterpart co-spatial with the line emission dislocated at 27 kpc from it in the source
plane. Moreover, for the physical separations between the individial sources with confirmed reshift, 
\cite{van19} suggest values of several tens of  kpc. These values are qualitatively in agreement with our results.  
The reconstruction of the morphology and position of the star-forming knots on the source plane required
deep MUSE observations \citep{van21a} and is still in progress. 
Further analysis of the datasets will be presented in a forthcoming work (Mestric et al., in preparation) and
will foster a more quantitative comparison between the D1+T1 complex and simulations.

\subsection{Discussion}

Despite the high densities resolved in our simulation due to the adopted
maximum resolution, the stellar aggregates do not reach extreme values in terms of density. 
This is in disagreement with what found in other studies. 

By means of a smoothed particle hydrodynamics code, \cite{kim18} used cosmological simulations with
minimum force resolution 1.4 $h^{-1}$ pc to study the formation of bound clusters in a halo of virial mass
$\sim~10^{10}~\rm M_{\odot}$ down to $z\sim 5$. 
The adopted star formation threshold is of $500$ cm$^3$ and the simulation includes feedback from radiation pressure,
stellar winds and energy and momentum deposition from SNe, all IMF-averaged and in a model with a \cite{kro01} IMF. 
They found that most stars are in loosely bound associations characterised by surface density $\sim 1 \rm M_{\odot}$pc$^{-2}$.
They interpret this diffuse component as composed by associations 
that inherit the properties of the parent molecular clouds, only marginally gravitationally
bound after turning a few percent of their mass into stars. A smaller fraction of the stellar mass is represented by
self-gravitating bound clusters that survive for long times (they are still present at the end of the simulation)
and with surface density $\sim 10^3 \rm M_{\odot}$pc$^{-2}$.
Although these systems represent a minority, they
have masses $10^6 \rm M_{\odot}$ and sizes of $\sim 10$ pc, compatible with local star clusters \citep{kru19}. \\
In a subsequent study and with a similar setup, \cite{ma20} found a larger population of these dense 
systems in a series of simulations targeting DM halos with virial masses between $\sim 10^{10}~ \rm M_{\odot}$ and
$\sim 10^{12}~ \rm M_{\odot}$. They found a significant population of bound clusters with masses
between $10^{3.5}~ \rm M_{\odot}$ and $10^{8.5}~ \rm M_{\odot}$ and half-mass radii typically between 6 pc and 40 pc.

In the work of \cite{kim16}, performed with RAMSES and characterised by comparable resolution
(although the simulations are carried on until $z~\sim~10$ and without modelling feedback from single stars), 
very compact systems (with size $\le 1$ pc) with typical
mass $\sim~6~\times~10^5~\rm M_{\odot}$ were found, therefore with densities comparable to the ones of local globular clusters. 
This work adopts a density threshold for star formation of $10^5$ cm$^3$ and different feedback prescriptions than in our work. 
These include (1) a pre-SN stellar feedback in the form of ionizing photons from massive stars, heating the cold gas to 
$2 \times 10^4$ K, (2) deposition of momentum (with no shutdown of radiative cooling)
in the cells occupied by the newly formed stellar particles and (3) SN explosions, 
in a model with a \cite{kro01} IMF. 
The pre-SN stellar feedback is weaker than the one adopted here (i.e. the 'delayed cooling'). 
In fact, as discussed in \cite{ros17}, by neglecting radiative cooling even for short times,
delayed cooling results in overefficient feedback with respect to other schemes. 
While in our model, due to the strong feedback, the collapse of the gas clouds is arrested at the onset of star formation,
the scheme adopted by \cite{kim16} allows for further collapse of the cold clouds. This contributes to attain high stellar
densities at the onset of SN explosions, 
which represent the events eventually arresting the star formation episodes.
Due to the similar setup and resolution but different physical ingredients, the simulation of \cite{kim16} represents the best 
case for a fair comparison with the one presented in this work. In the future, a deeper investigation of the main parameters regulating
star formation and feedback will be performed, in order 
to investigate further in which conditions the dominant star formation mode occurs in dense and loosely bound stellar aggregates.

\subsubsection{Size-Mass relation}

\begin{figure*}
\includegraphics[width=12.cm,height=12.5cm]{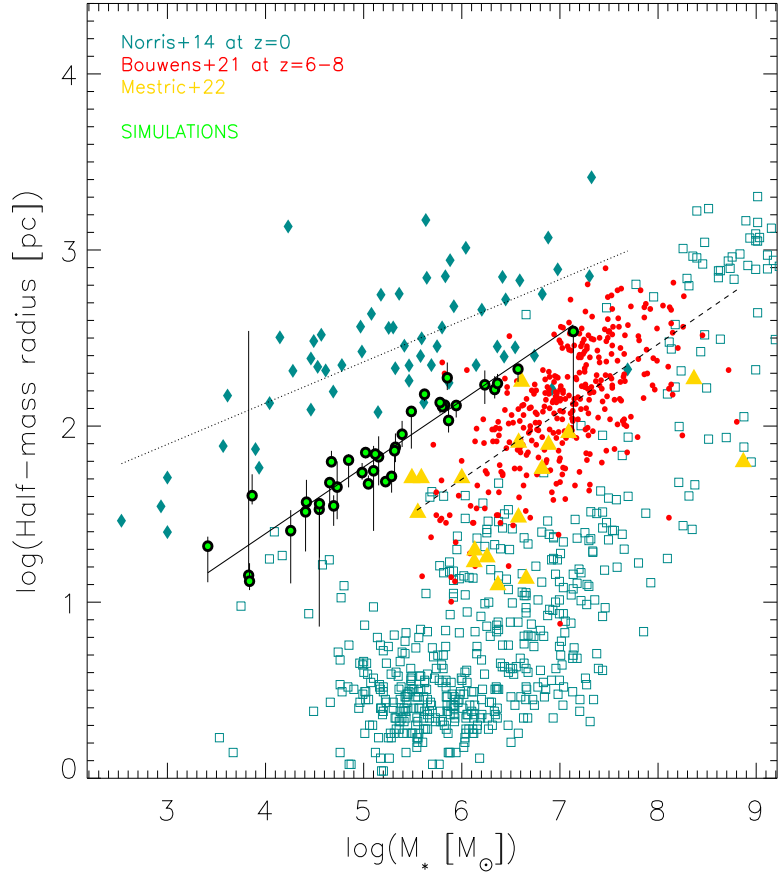}
    \caption{Relation between stellar mass and size for the stellar clumps identified in our high-resolution simulation
    (solid green circles) compared to observational datasets from the literature. For each clump, 
    its size is computed from the half-mass radius of its stellar surface density profile, considering three different viewing angles, i. e.
    along the x-,y- and z-axis. The green circles represents the median values calculated considering the three projections,  
    whereas, for eah clump, the error bar is computed from the distance between the minimum and the maximum size.      
    The other symbols are observational values 
    from a sample of lensed faint galaxies with photometric redshifts in the range $6 \le z \le 8$ (\citealt{bou21}, red solid circles),
    a set of spectroscopically confirmed lensed clumps at $z\gtrsim6$ (\citealt{mes22}, yellow triangles) 
    and from an extended local sample which includes globular clusters, spheroids, dwarf ellipticals, ultra-compact dwarfs and
    dwarf spheroidals (\citealt{nor14}, dark cyan symbols). The dark cyan solid diamonds represent local dwarf spheroidals, whereas the open
    squares indicate all the other local systems. 
    The solid, dotted and dashed lines represent linear fits (expressed by $y=a+b \cdot x$) 
    to the relations of the simulated
    clumps (with $a=-0.12$, $b=0.38$) the dSphs of \citet{nor14} (with $a=1.19$, $b=0.23$) and the lensed systems of \citet{bou21} ($a=-0.6$, $b=0.38$), respectively.} 
    \label{size_mass_fig}
\end{figure*}

In order to gain further insight on the properties of our clumps
compared to observations and analogs in the nearby Universe, we examine one fundamental 
scaling relation between two structural parameters of these systems, i. e. the size vs mass relation, shown in Fig. ~\ref{size_mass_fig}.
The sizes of the simulated clumps are compared with observational datasets from the literature.

\cite{bou21} presented a sample of faint galaxies with photometric redshifts in the range $6 \le z \le 8$ magnified by the
Hubble Frontier Fields clusters. 
Their sample includes 330 galaxies with size and mass measurements obtained by means of
various public models for the gravitational lenses. Their objects have stellar mass in the range
$10^5 \rm M_{\odot} \le M_{*} \le 10^8 \rm M_{\odot}$ and effective radii $R_{\rm e}$,
in the range $10$ pc$\le R_{\rm e} \le$ 1000 pc. 
The data of \cite{bou21} (red circles in Fig. ~\ref{size_mass_fig}) build a clear, positive correlation in the size-mass plot. 
Due to their extended ranges of size and mass, the lensed sources are broadly classified in the cathegory of
'star cluster complexes', similar to the local example of 30 Doradus,
an extended HII complex with an estimated mass of $10^6 \rm M_{\odot}$ and size of 100 pc located in the Large Magellanic Cloud \citep{leb08}.
This classification is performed in order to distinguish them from regular star clusters, whose size-mass relation
is much debated, with a nearly flat behaviour as found in the LEGUS sample of local dwarf and spiral star-forming galaxies
(\citealt{ryo18}, see also \citealt{kru19}). 
One possible interpretation for the extended sizes of the lensed sources is that they might comprise groups of star clusters, 
appearing in observed fields as blended together into one single object, although it is not possible to exclude
that at high redshift, star clusters, and stellar aggregates in general,  might follow a size-mass relation different than the local one. 
In Fig. ~\ref{size_mass_fig}, we also show a subset of clumps at $z\gtrsim6$ from \cite{mes22} and 
a compilation of local objects collected by \cite{nor14}, which include
a variety of systems, i.e., besides globular cluster, it comprises spheroids, dwarf ellipticals, ultra-compact dwarfs and
dwarf spheroidals (see \citealt{nor14} for further details on the compilation).

For the calculation of the size of the simulated clumps, we have considered 
three different projections, i. e.  along the x-, y- and z-axis. 
For each clump, the green circle represents the median value of the three projections,  
whereas the error bar is computed from the distance between the minimum and the maximum size. 

The simulated clumps present sizes significantly larger and a flatter size-mass relation than the lensed sources,
building an intermediate sequence between high-$z$ clumps and local dSph galaxies.

Some clumps present non-spherical shapes, in particular in the central region, where
tidal distorsions are frequent, as visible also from the extended distributions of stars (see the bottom panel of
Fig.~\ref{fig_sig}). 
From a preliminary analysis performed assuming non-spherical radial bins, the deviations from sphericity of the clumps 
has negligible impact on the size-mass relation. 
The flattening of the clumps will be analysed in detail in a forthcoming work (Pascale et al., in prep.).


It is worth stressing that the bridging of dense clumps represents a major
downside of friends-of-friends approaches with a potential non-negligible impact on the size-mass relation, since some  
groups can be intepreted as very extended clumps rather than multiple, more compact clumps. 
In a forthcoming paper, we will address this issue in detail, with special care on the central region, where 
we will test the robustness of our results against the adopted linking length and 
compare the results obtained with fof with 
other clump identification tools (Pascale et al., in prep.).  

Moreover, the clump segmentation has been performed in the three-dimensional space,
whereas the sizes have been calcualted as half-mass radii, considering 2D projections, which might 
seem inconsistent.
However, for the chosen viewing angles and eliminating the central region
from the clump identification, the clumps never overlap in projection, so we expect fof and other clustering tools to identify
the same number of clumps.

In the future, we will investigate more in detail the effects 
of stellar feedback and other baryonic processes on the slope of the size-mass relation.  
A more complete study of the main scaling relations of the simulated clumps and a more thorough
comparison with extended observational datasets of lensed clumps \citep{mes22}  
(including also  the calculation of the surface brightness of the symulated systems)  
will be presented in a forthcoming work. 


\section{Summary and Conclusions}
The aim of this work is to present the first results
obtained with a new set of zoom-in cosmological simulations with unprecedented features.
To our knowledge, our cosmological simulations are among the first ever performed 
with a sub-pc resolution, carried on down to $z\sim 6.1$ (i.e. for a cosmic time interval of 0.9 Gyr) and
including the feedback of individual stars. 
The model is originally designed to model the spectroscopically-confirmed, lensed star-forming complex D1+T1 
observed at $z=6.14$ (\citealt{van19,cal21}), which includes several clumps, distributed across a few 10 kpc - wide region.
One of the detected star-forming knots presents a very dense region at its centre, with density
fully compatible with the values expected in a proto-globular cluster.
The aim to probe substructures within the first, compact clumps represents the main motivation
for adopting a sub-pc resolution.

The target DM halo is characterised by a maximum mass resolution of 200 $\rm M_{\odot}$ and $2\times10^8$ collisionless particles. 
The simulations include star formation and can track the energy, mass and metals deposited by individual massive stars and AGB 
with mass $\ge 3~\rm M_{\odot}$.
The generation of single stars occurs via stochastic, direct IMF sampling, following the method presented
in \cite{sor17} (see also \citealt{and20}), whereas low-mass (i.e. with mass $<3~\rm M_{\odot}$) stars are stored into standard stellar particles.
This choice was performed to limit the total number of stellar particles. 
As for massive stars, we take into account both pre-SN and SN feeback.
In both cases, these sources deposit instantly variable amounts of mass and energy, proportional
to their initial mass.
Our model represents a simplified 
feedback scheme designed to account for the effects of pre-SN and SN feedback.  
We adopt a 'delayed cooling' feedback scheme similar to \cite{tey13} to prevent artificial
radiative loss of the energy injected by the stars. In this scheme, 
radiative cooling is temporarily switched off in suitable cells where the feedback
is stored also in a passive tracer, advected with the flow and aimed to represent an unresolved, 'non-thermal' energy component. 
The native delayed cooling implementation \citep{tey13} was created to model stellar particles in lower resolution runs and is unsuited to our case, where the non-thermal 
energy deposited in a single cell can be very small and, at the same time, the gas can be very dense.
We propose a simple, versatile recipe to overcome this issue, in which the flag used for switching off radiative cooling
is posed equal to its maximum value as soon as any star releases energy, and relying on the standard implementation to 
account for the fast dissipation of the non-thermal energy.
At our maximum 
resolution, our feedback implementation is efficient in regulating
star-formation, but in the future, it needs to be improved in
various aspects, which include a more gradual release of energy in the pre-SN phase in the form of stellar winds, ionization 
and injection of momentum. 

We hope that the proposed approach may be useful in future studies including individual stellar feedback in grid codes. 
Our main results can be summarised as follows.
\begin{itemize} 
\item The onset of star formation occurs
in dense ($\Sigma_{\rm gas}>10^2 \rm M_{\odot}/pc^2$) gas clumps lying at the centres of networks of
low-pressure, cold (with typical temperatures of $\sim 10^3$ K) and elongated 
filaments with surface densities $\sim 1 \rm M_{\odot}/pc^2$.  
The sites where star formation ignites 
are characterised by maximum densities of the order of $10^5$ cm$^{-3}$ and 
pressure values  P$_{\rm grav}$/k  $>  10^7$ K/cm$^3$, corresponding to the pressures of the local, turbulent regions 
where the densest stellar aggregates form \citep{elm97}. 
In our highest resolution run, the thin, pc-wide dense shells of the energetic
bubbles generated by a handful of individual massive stars can be resolved.
Our feedback scheme correctly accounts for the SN-driven, hot gas with characteristic temperature $>10^6$ K. 
\item Due to the high resolution, our simulation allows us to resolve stellar populations with
age differences smaller than 1 Myr.
The first stars originate in scattered, loosely-bound aggregates.  
As cosmic time evolves, more and more stellar clumps form at the centre of the DM halos
with mass density $>10^2 \rm M_{\odot}/pc^2$ and larger and larger volumes 
are enriched with metals due to exploding SNe. Typically, the maximum metallicity is $Z \sim 10^{-3}$, reached in regions
where the largest stellar clumps (of mass $\sim~10^6~\rm M_{\odot}$) are located. 
The total SFR increases with cosmic time, to reach the value of $0.1~\rm M_{\odot}$yr$^{-1}$ at the endpoint of
our simulation ($z=6.14$, corresponding to a cosmic time of 0.92 Gyr). 
\item The total stellar mass at $z=6.14$ is 3.4$\times10^7~\rm M_{\odot}$,
in satisfactory agreement with the observed stellar mass of the D1+T1 system (2.3$\times10^7~\rm M_{\odot}$)
and considering that other fainter star-forming knots at the same redshift are known to
be part of the observed star-forming complex \citep{van19}. 
The simulated highest mass clumps have sizes (as measured from their stellar surface profiles) of $\sim 100$ pc.
On the other hand, the measured half-light radius of D1 is 44 pc.
The average stellar density
of the simulated clumps is therefore lower  by $\sim 1$ order of magnitudes than the observed ones.
No simulated clump shows any presence of signficantly denser stellar sub-structures, whereas
a very dense stellar system, (characterised by mass $~10^6~\rm M_{\odot}$ and radius
$<13$ pc) is reported at the centre of D1 \citep{van19}.
The total SFR of the simulated system is more than two orders of magnitude lower than the observed
best value  ($15.8~\rm M_{\odot}/$yr, \citealt{van19}), which is however affected by several systematic uncertainties related
to the magnification and fitting procedure.
Were the SFR of D1 of the same order of magnitude as the total SFR of our simulations, this would 
alleviate the significant deviation of D1 from the observed SFR-$L_{\rm CII}$ relation \citep{car18}
reported by \cite{cal21}. 
\item The study of the size-mass relation is useful to compare 
the properties of our clumps with possible observational analogues at both high redshift and in the local Universe.
We have calculated the size of the simulated clumps
as half-mass radii, considering 2D mass distributions and considering three different projections, i. e. along the x-,
y- and z-axis. 
We compared our results with the ones from 
samples of faint galaxies with photometric redshifts in the range $6 \le z \le 8$ magnified by the
Hubble Frontier Fields clusters \citep{bou21} and a  compilation of local objects which include 
a variety of systems (GCs, spheroids and dwarf galaxies of various types, \citealt{nor14}).
The simulated clumps present larger sizes than the ones of the lensed clumps, 
building an intermediate sequence (in both terms of slope and normalization) between high-redshift systems and 
local dSph galaxies.
In a forthcoming paper, we will address how some issues
of clumps idetification and segmentation (such as bridging of dense clumps)  
affect their properties, comparing the results obtained with fof  
with other clump identification tools (Pascale et al., in prep.).  
\end{itemize}
On the observational side, efforts are in progress to 
collect larger and larger 
catalogues of high-redshift clumps where various parameters can be measured,
including mass, size and SFR \citep{mes22}. 
These studies will be fundamental drivers for 
our future work, which will focus on a more complete study of the main scaling relations of the simulated clumps and a more thorough 
comparison with the observational datasets. 
Stellar feedback plays a key role in the formation and evolution of stellar structures.
Other simulations with different stellar feedback see the early formation of dense stellar
clusters at the centre of DM halos (\citealt{kim16},\citealt{kim18}). 
Mass loss from stellar winds and supernova feedback causes also a significant expansion of young stellar clusters
(e. g., \citealt{bau08,ma20,ves21,sol21}).
The role of these processes and those of the early dynamical evolution need to be investigated in detail,
along with with their effects on the structural properties and the scaling relations of compact stellar systems.

\section*{Data Availability}
The data underlying this article are available from the 
corresponding author, upon reasonable request.

\section*{Acknowledgements}
We are grateful to an anonymous referee for several useful suggestions. 
We acknowledge support from PRIN INAF 1.05.01.85.01 and INAF Main-Stream 1.05.01.86.31. 
AL acknowledges funding from MIUR under the grant PRIN 2017-MB8AEZ.
EVa, MM and PR acknowledge support from PRIN-MIUR 2017WSCC32. 
EVe acknowledges support from NSF grant AST-2009193. 
This research was supported in part by Lilly Endowment, Inc., through its support for the Indiana University Pervasive Technology Institute.







\appendix

\section{Properties of other stellar clumps}

\begin{figure}
	\includegraphics[width=8.cm,height=11cm]{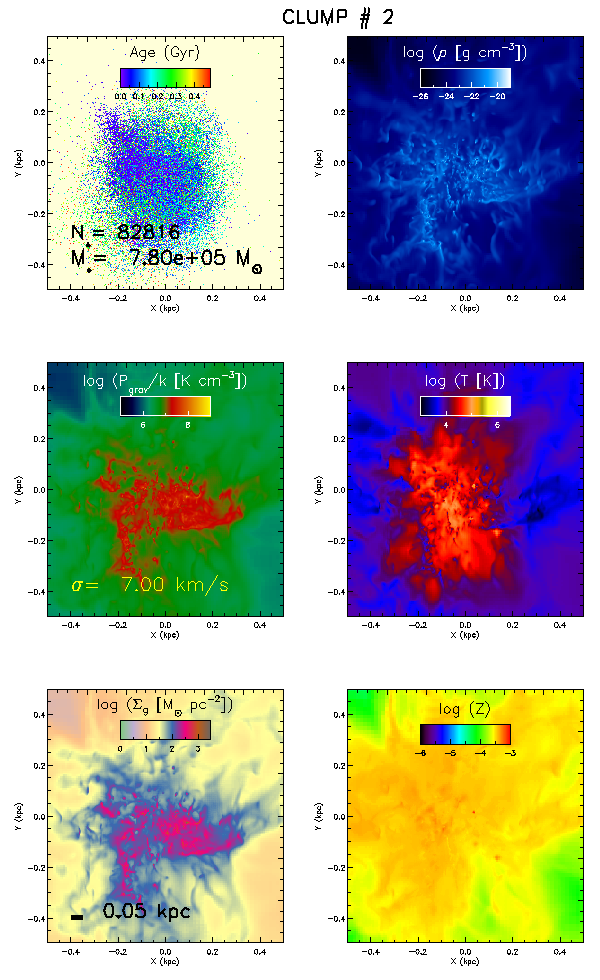}
\caption{Main properties of clump  2 of Fig. \ref{fig_sig} and slice maps of various properties of the gas, each one
centered on the centre of mass of the clump. Panels as in Fig. \ref{fig_clump1}.}
    \label{fig_clump2}
\end{figure}

\begin{figure}
	\includegraphics[width=8.cm,height=11cm]{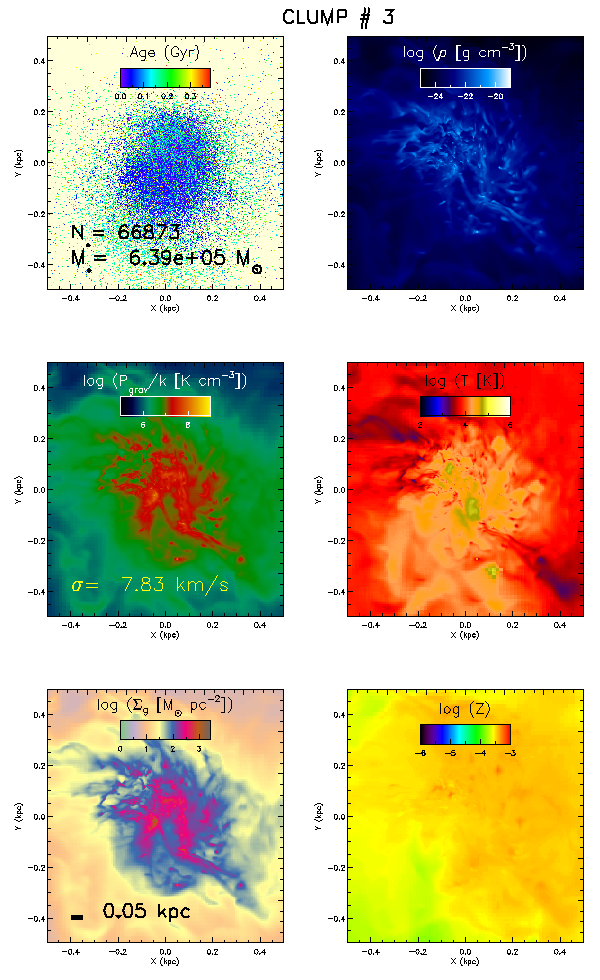}
\caption{Main properties of clump  3 of Fig. \ref{fig_sig} and slice maps of various properties of the gas, each one
centered on the centre of mass of the clump. Panels as in Fig. \ref{fig_clump1}.}
    \label{fig_clump3}
\end{figure}

\begin{figure}
	\includegraphics[width=8.cm,height=11cm]{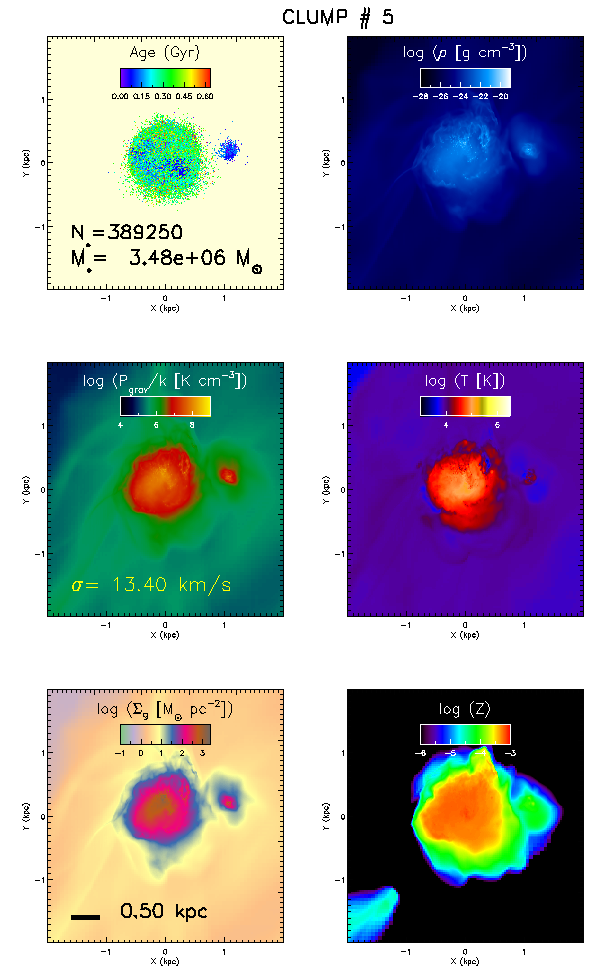}
\caption{Main properties of clump 5 of Fig. \ref{fig_sig} and slice maps of various properties of the gas, each one
centered on the centre of mass of the clump. Panels as in Fig. \ref{fig_clump1}.}
    \label{fig_clump5}
\end{figure}

In Figures ~\ref{fig_clump2}, ~\ref{fig_clump3} and ~\ref{fig_clump5}
we present other cases of stellar clumps identified with fof and the properties of the gas surrounding them.
Clumps 2 and 3 have similar mass and overall structure, but occupy different positions in the box, as clump 2
is near the centre and clump 3 is 10 kpc away from it. It is interesting to compare their properties,
remarkably similar in terms of stellar age, gas density, pressure and metallicity despite the different
environment surrounding them.
Clump 5 is the most massive ($M_{*}=3.48 \times 10^6 \rm M_{\odot}$) and has older stars than the other cases and 
a significantly smaller structure (with mass $M_{*}\sim 2  \times 10^4 \rm M_{\odot}$) besides,
visibile on the right side of Fig. ~\ref{fig_clump5}.



\section{Phase diagrams}
\begin{figure*}
	\includegraphics[width=8cm,height=9cm]{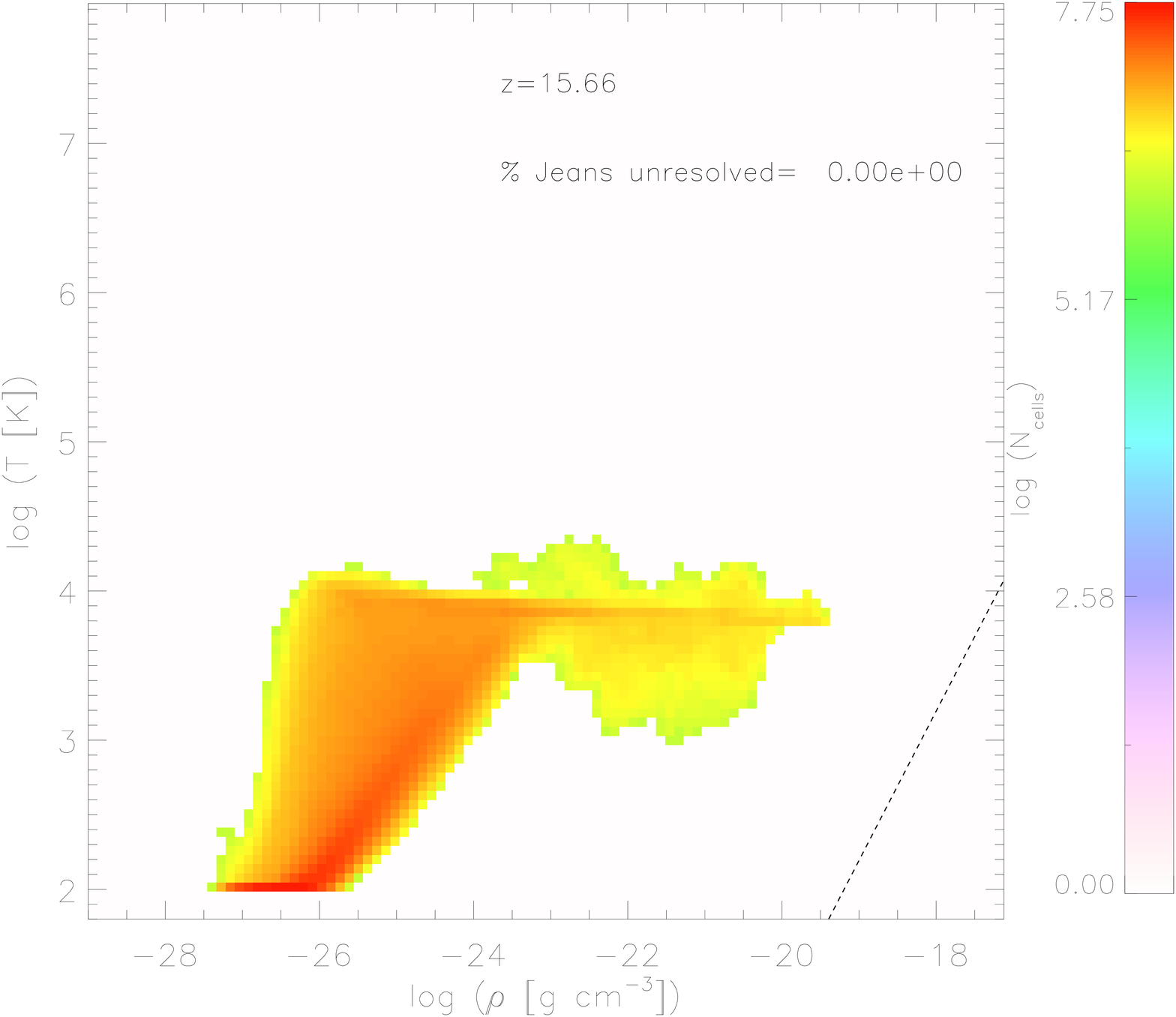}
	\includegraphics[width=8cm,height=9cm]{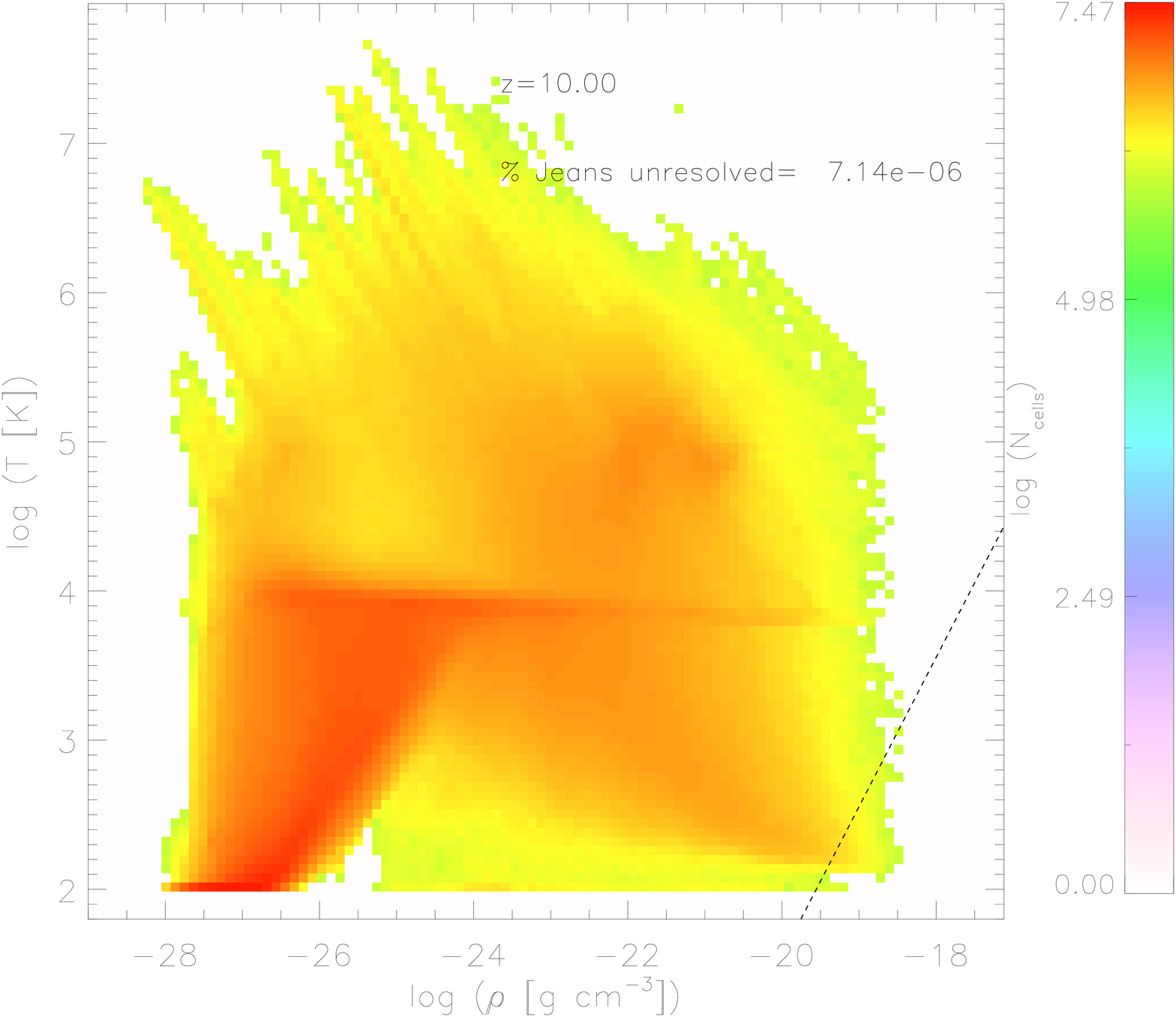}

\caption{Temperature–density phase diagrams at $z=15.66$ (left) and $z=10$ (right) in our highest resolution simulation.
The color scale indicates the number of cells. 
The diagonal black dashed lines  indicate the analytic relation between T and $\rho$ obtained by requiring that
the Jeans length is resolved by at least 4 cells. In each panel, the fraction of cells where the Jeans length is unresolved
(i. e. where $\lambda_{\rm J} < 4 \Delta x$) is reported in the top-right corner.}
    \label{fig_phase}
\end{figure*}

In Fig~\ref{fig_phase} we show two phase diagrams (i.e. gas temperature vs density) calculated at two
different redshifts. These plots are useful to address the issue of artificial fragmentation, a problem
common to star formation simulations, where the cold gas may reach particularly high densities. 
Gas simulations where the Jeans length, defined as
\begin{equation}
\lambda_{J}=\left(\frac{\pi c_s^2}{G \rho}\right)^{1/2}
\end{equation}
(where $\rm c_s$ is the sound speed), is under-resolved experience the
phenomenon of artificial fragmentation, in which perturbations
arising from the discrete nature 
of the equations of self-gravitational hydrodynamics can grow into non-physical
clumps. 
\cite{tru97} demonstrated that the condition in which artificial clumps are avoided
is when $\rm \lambda_{\rm J} \ge 4 \Delta x$, where $\Delta x$ is the minimum cell width.
The black diagonal lines in Fig. ~\ref{fig_phase} represent
the analytic relation between T and $\rho$ obtained by requiring that
the Jeans length is resolved by at least 4 cells
\begin{equation}
T_{Jeans} \ge \frac{\rho}{m_H} \left(\frac{4 \Delta x}{16~pc}\right)^2.
\end{equation}
The fraction of cells where the Jeans length is not resolved 
is reported in both panels of Fig.~\ref{fig_phase}. 
At $z=6.14$, this fraction is  $3 \times 10^{-6}$.

\section{Stellar IMF}
\begin{figure*}
	\includegraphics[width=10cm,height=10cm]{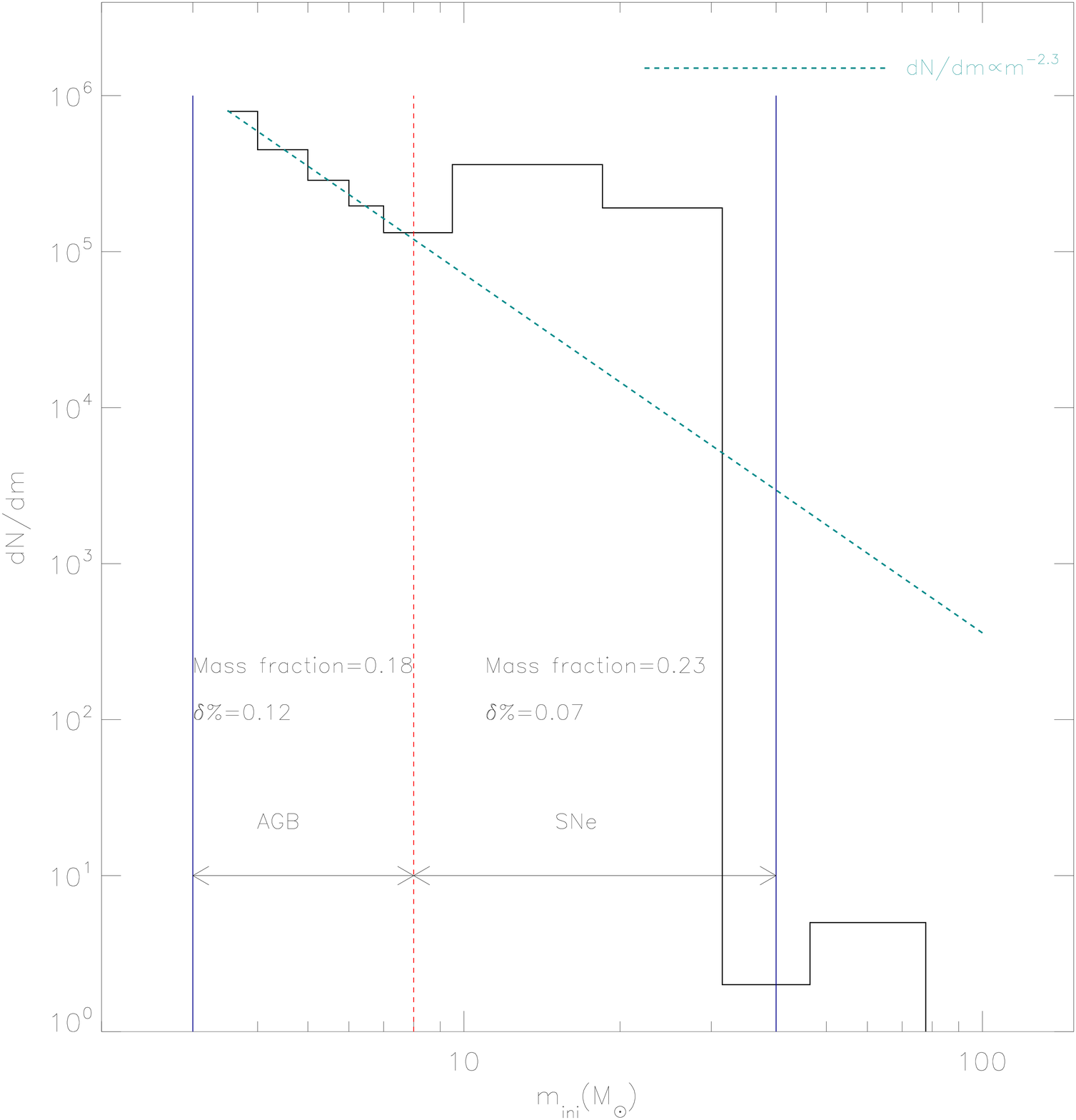}
\caption{Initial mass function of individual stars obtained in our simulation (black histogram),
compared with the one of \citet{kro01} in the same mass range (dark-cyan dashed line). 
The range in which individual stars are created
 is between 3 $\rm M_{\odot}$ and 100 $\rm M_{\odot}$
The vertical red dashed line marks the separation mass between
AGB and the progenitors of core-collapse SNe, i.e. the massive stars with  initial  mass  $8~\rm M_{\odot}< m_{\rm ini} <40~\rm M_{\odot}$.
The vertical solid blue lines enclose the AGB+SNe mass range. 
We report the total mass fractions of individual AGB and SNe, along with the deviation $\delta\%$ from the value obtained from the integral
of the \citet{kro01} in the corresponding mass range.}
\label{fig_imf}
\end{figure*}

In our simulations we follow stellar particles,
representing populations of low-mass stars (i. e. with initial mass $<3$ M$_{\odot}$) and
individual stars, with mass $>3$ M$_{\odot}$.
In Fig.~\ref{fig_imf} we show the initial mass function of individual stars obtained in our simulation,
compared with the one of \cite{kro01} in the same mass range. 
The IMF is slightly top-heavier than the \cite{kro01}, with an actual number of MS $\sim 1.2$ times the one
expected with the base IMF.
The mass fractions in the AGB and SNe ranges are reported, along with the deviations from the base IMF.
In both ranges, deviations from the  \cite{kro01} IMF are close to 10 \%.

\bsp	
\label{lastpage}
\end{document}